\documentstyle[aps,pre,epsfig]{revtex}

\begin{document}

\draft

\title{Frictional mechanics of wet granular material}
\author{Jean-Christophe G\'eminard$^{1 ,2}$, Wolfgang Losert$^{1}$ and Jerry 
P. Gollub$^{1,3}$}

\address{$^{1}$ Physics Department, Haverford College, Haverford, PA 19041}

\address{$^{2}$ Laboratoire de Physique de l'E.N.S. de Lyon, 46 All\'ee 
d'Italie, 
69364 Lyon Cedex,France}

\address{$^{3}$ Physics Department, University of Pennsylvania,
Philadelphia PA 19104}
 
\date{\today}

\maketitle

\begin{abstract}

The mechanical response of a wet granular layer to imposed shear
is studied experimentally at low applied normal stress. The granular material
is immersed in water and the shear is applied by sliding
a plate resting on the upper surface of the layer.
We monitor simultaneously the horizontal and the
vertical displacements of the plate to submicron accuracy with millisecond 
time resolution. The relations between the plate displacement,
the dilation of the layer and the measured frictional 
force are analyzed in detail.  When slip begins, the dilation increases 
exponentially over a slip distance comparable to the particle radius.
 We find that the total dilation and the steady state 
frictional force do not depend on the driving velocity, 
but do depend linearly on the applied normal stress.  
The frictional force also depends linearly on the dilation rate (rather than the 
dilation itself), and reaches a maximum value during the transient 
acceleration.   
We find that the layer can temporarily sustain a shear stress that is 
in excess of the critical value that will eventually lead to slip.  
We describe an empirical model that describes much of what we observe.  
This model differs in some respects from those used previously 
at stresses $10^{6}$ times larger.

\end{abstract}           


\section{INTRODUCTION}

The response of any material to an applied shear stress is an important 
mechanical property.  We are concerned here with granular materials, 
for which the response to shear stress amounts to friction.  
We have previously reported an extensive study of dry granular friction 
under low applied normal stress \cite{Gollub,Nasuno}, and here we extend 
that work to the case of wet materials.  Although our motivation is primarily 
fundamental, it should be recognized that a study of granular friction, 
including the wet case, is of potential interest in connection with 
applications such as the processing of powders. 
It may also be of conceptual or heuristic interest in connection with 
seismic phenomena, though the applied stresses in our work are lower by 
a factor of $10^6$.  Do the friction laws obtained at pressures of 
100 MPa extrapolate to such low stresses, or do completely new effects occur? 
An excellent review of geophysically inspired experiments conducted at 
high pressures has been given by Marone \cite{Marone}.  
Much of the work described in that review was concerned with the 
determination of friction laws that incorporate memory effects.  
These state variable friction laws are also reviewed in the book by 
Scholz \cite{Scholz}, and a recent review of friction that includes 
consideration of microscopic physics has been given by Persson 
\cite{Persson}. 

We have several different motivations in considering the case of a granular 
medium that is immersed in a fluid.  It is well known that even slight 
amounts of water adsorbed on a granular medium (or a rough surface) can 
affect friction.  In one recent paper, the condensation of small amounts 
of water near points of contact was shown to produce substantial 
strengthening of a granular material \cite{Ciliberto}.  
To eliminate this effect, one can fill the pore spaces entirely with a fluid, 
i.e., submerge the material.   Once we undertook such an experiment, 
we found that immersion has additional merits :  Lubrication reduces 
the friction so that the transition from stick-slip dynamics to continuous 
sliding occurs at very low velocity.  Thus, one can study the dynamics of 
the transition from rest to steady sliding over a wide range of velocities, 
which turns out to allow delicate tests of
friction laws.  

Broad references to the literature on granular friction were given in the 
introduction to our previous paper \cite{Gollub}, and we do not repeat that 
material here. Works specifically concerned with friction in wet granular 
materials appear to be rather limited. However, it is important to mention 
the experimental work of Marone {\it et al.} on the frictional behavior of 
simulated fault gouge \cite{Marone90}. They studied the response of immersed 
granular material subjected to shear deformation and observed both a 
noticeable 
dilation of the material and a consequent velocity strengthening. These 
results were obtained at large normal stress (50-190 MPa) in sand. 
The work presented here differs from their work in two  essential respects: 
First, the normal stress applied to the layer is at least six orders of 
magnitude smaller in our work (typically 20 Pa). Second, the granular 
material considered here consists of spherical glass beads with a narrow 
size distribution. 
We point out in a later section the similiarities and differences in
the resulting frictional dynamics.  A thorough 
controlled study of the frictional behavior of an immersed granular material 
under low stress, comparable to what we have presented previously for dry 
materials, has not appeared previously to our knowledge.  

The experimental setup is similar to that described in 
Refs.~\cite{Gollub,Nasuno,Perrin,Baumberger}. 
The sensitivity achieved in the present paper is an advance over previous 
work.  We are able to measure both 
vertical and horizontal displacements to a precision of about 
$0.1~\mu \rm m$, and are able to do so with excellent time resolution
of $<0.1~{\rm ms}$.  
It is true but surprising that precision for displacement measurements 
much better than the particle size is necessary in order to explore 
the dynamics fully.  We focus attention here particularly on the vertical 
dilation that accompanies slip, since it is clear that dilation is a 
necessary dynamical variable in any granular friction law.

The present article is organized as follows. Our experimental methods 
are described in Sec.~II and the observations are presented in detail 
in Sec.~III. In Sec.~IV we compare our results to an empirical model 
and previous work and then we conclude.

\section{EXPERIMENTAL METHODS}
\subsection{Introduction}

\begin{figure}
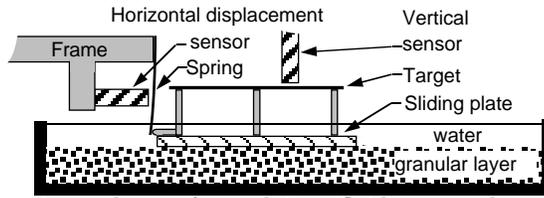

\caption{Schematic diagram of the experimental setup for studying a 
fluid-saturated
granular layer with sensitive measurement of horizontal and vertical 
positions.}
\label{setup}
\end{figure}

A schematic diagram of the experimental setup is shown in Fig.~\ref{setup}.
The granular material resides in a rectangular tray whose lower surface has 
been coated permanently with a layer of particles to ensure that no sliding 
occurs at this interface when the layer is sheared.  
A given amount of granular material is then spread out on the tray
and it is filled with distilled water.
A homogeneous layer of material 3 mm deep is obtained by sliding a 
straight rod along the side edges of the tray.  The sliding plate is gently 
placed on top of the granular layer, additional water is added to ensure 
that it is submerged and covered by about $1~\rm mm$ of water, and a visual 
check is made to ensure that no air
bubble has been trapped underneath the plate.

Shear stress is imposed on the granular layer
by translating the plate over its upper surface.
In order to transmit the shear stress, the bottom surface of the sliding
plate is roughened in one of two
ways: (a) If visualization is not required, a glass plate is used with a 
monolayer of granular material glued to its lower surface.  (b) When 
visualization is desired, a transparent acrylic plate is used; its lower 
surface is ruled with parallel grooves 2 mm apart and about 0.1 mm deep to 
pin the material.  Most of the experiments reported here have been done 
using (b), but we have checked that (a) gives essentially the same results.

The dimensions of the sliding plate are length $L = 8.15~\rm cm$,
width $W = 5.28~\rm cm$, and thickness $T = 0.88~\rm cm$;  the direction of 
travel is along the length.  The in-plane dimensions of the tray 
(i.e. $11 \times 18.5~\rm cm^2$)
are much larger than those of the sliding plate; no boundary
effect has been observed. 

The granular material used in most of the experiments reported here consists 
of spherical glass beads with a mean diameter of $103 ~\mu \rm m$ and a 
standard 
deviation that is $14 ~\mu \rm m$ from Jaygo, Inc.  We have checked that our 
conclusions remain valid for other samples with mean diameters of 
about $200$ and $500$ $\mu\rm m$.

Whether or not a layer of beads was glued to the bottom of the container 
did not affect the experimental results described in this paper. 
We conclude that the shear zone is smaller than the layer height 
in all cases considered. 
The small total vertical dilation of
the granular layer during shear, described in Sec. III, 
indicates that the shear zone height is, at most, a few beads. 
The size of the shear zone can become comparable 
to the height of the granular layer for the largest beads 
(for which the layer height is approximately six bead diameters) 
at low pulling velocities. However, such a case is not included in 
the experimental results.

Visual inspection of the top layer of beads underneath the transparent
acrylic plate shows that the beads are arranged randomly. The arrangement
remains random after repeated sliding of the plate. The organization 
of beads in deeper layers cannot be determined visually. 

 The volume fraction of
glass beads is $0.63 \pm 0.02$ for both wet and dry cases.
We find that shaking and pressing of the material does not 
increase the volume fraction measurably.
The measured value is close to the numerically obained estimate of $0.63$ for 
random close packing of uniform spheres~\cite{Brady93}; 
a slightly larger value would be expected for random close packing
if there is some variation in bead size. 

\subsection{Driving system and frictional force measurement}

In order to control the stiffness of the driving system,
the plate is pushed with a blue tempered steel leaf spring
(constant $k = 189.5~\rm N/m$) connected to a translation stage.
The stage is driven at constant velocity $V$
(between $0.1~\mu \rm m/s$ and $1~\rm mm/s$) by a computer-controlled
stepping-motor which turns a precision micrometer. The coupling
between the spring and the plate is accomplished through a rounded
tip which is glued to one end of the plate.
This makes it possible for the plate to move both horizontally and
vertically as the thickness of the granular layer varies during
the motion.

The elastic coupling through the spring allows relative motion of the plate
with respect to the translator. The horizontal relative motion is monitored 
by measuring the displacement $d(t)$ of the spring from its
rest position at the coupling point with an inductive sensor, model EMD1050
from Electro Corporation, as in Refs.\cite{Gollub,Perrin}. 
A typical trace of $d$ as a function of time is shown in 
Fig.~\ref{illustration}(a).

The differential equation governing the position $x(t)$ 
of the sliding plate in the laboratory frame reads
\begin{equation}
M {d^2x \over dt^2} = k (V t - x) - F = k d - F ,
\label{diff}
\end{equation}
where $M$ is the mass of the sliding plate, $k$ is the spring constant,
and $V$ is the pulling velocity. Here $F$ denotes the frictional force,
which depends on time. For the parameter range used in the 
experiment, the inertia of the sliding plate can be neglected and
the frictional force is then $F = k d$.
We estimate that the displacement $d(t)$ is known within $0.1~\mu \rm m$
so that the frictional force $F$ is measured within $2 \times 10^{-5}~\rm N$.  
This is typically about $0.01\%$ of the normal stress.

\subsection{Vertical displacement measurement}

\begin{figure}
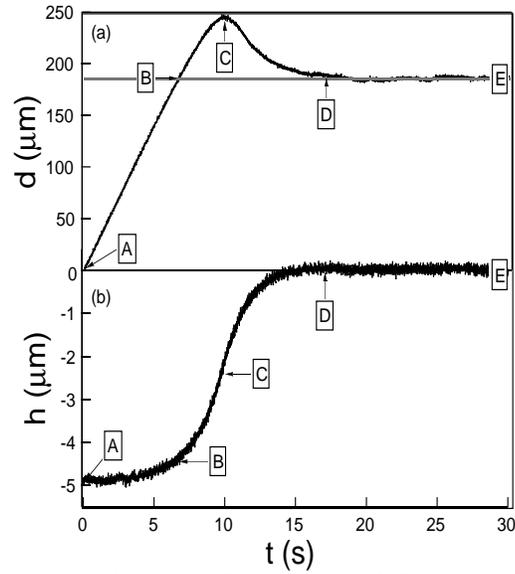

\caption{Typical behavior (a) of the spring displacement $d(t)$ and
(b) of the vertical position $h(t)$ as a function
of time $t$ ($k = 189.5~\rm N/m$, $M = 14.5~\rm g$,
$V = 28.17~ \mu \rm m/s$). [The different stages of the plate
motion, labelled A-E, are discussed in section III.C.1.]}
\label{illustration}
\end{figure}

The vertical displacement $h(t)$ of the plate is monitored
with a second inductive sensor as shown in Fig.~\ref{illustration}(b).
The target of the sensor is
a blue tempered steel plate connected to the sliding plate with
the help of five pillars (diameter = $3.4~\rm mm$). The target is not 
perfectly flat and parallel to the sliding plate; the measurements
must be corrected to exclude this source of error 
at the micron level of sensitivity.
Once this is done, we estimate that the vertical position of the plate
is known within about $0.2~\mu \rm m$. We also point out that $h(t)$ is
determined only to within an additive constant; We choose the constant
so that $h(t \to \infty) = 0 $ in the steady-state regime
[e.g. the right hand side of Fig.~\ref{illustration}(b)].

\subsection{Applied normal stress: effect of immersion}
  
The plate is immersed in water, so it experiences a buoyant 
force. The mass $m$ of the displaced volume of water
(volume of the plate + volume of $1~\rm mm$ length of the pillars) is
$m = (12.5 \pm 0.5)~\rm g$. In the following, we define the mass
of the sliding plate $M$ to be its real mass reduced by $m$. 
For instance, most of the results given below are obtained
with a sliding plate whose actual mass is $27~\rm g$, so that
$M= (14.5 \pm 0.5 )~\rm g$.
The mass of the plate must be small enough that the
plate does not sink significantly into the granular layer.
We estimate the
density of the water-glass beads mixture to be about $2~{\rm g/cm^3}$.
Then,  assuming that the granular layer behaves like a
dense fluid, estimating the equilibrium depth of the bottom
surface of the plate, and assuming that the plate sinks
(during translation) when this depth equals the plate thickness, we 
determine that the plate should be expected to sink 
into the granular bed for masses larger 
than $40~\rm g$.
Experimentally, we find that the granular layer can sustain
somewhat larger loads but we limited our study to $M < 40~\rm g$.
 
\section{EXPERIMENTAL RESULTS}

\subsection{Basic observations}

We always prepare the system for measurements in the same way to assure
reproducible experimental conditions.
The granular layer is first installed as previously described, and the 
plate is then pushed at constant velocity ($20~\mu \rm m/s$)
until a steady state regime is reached.  During this initial ``break-in'' 
phase,  the plate slightly sinks
in the granular layer and a small heap forms at the leading edge.
This heap has been proven to have a negligible
effect on the frictional force by turning the plate through 
90 degrees. The
results obtained by pushing the plate either along its length
or along its width differ only slightly, despite the factor 
of 1.5 difference in dimensions.
The spring is then pulled back until it loses contact with 
the rounded tip so that no stress is initially applied for the measurements 
of interest. 

\begin{figure}
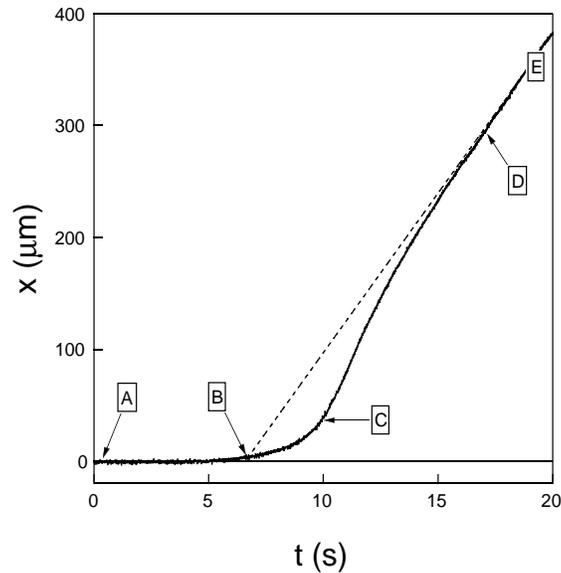

\caption{Position of the plate $x$ as a function of time
 $t$ ($k = 189.5~\rm N/m$, $M = 14.5~\rm g$,
$V = 28.17~ \mu \rm m/s$). The dashed line
indicates the asymptotic regime.}
\label{x_vs_t}
\end{figure}

Once the sample has been prepared as just described, the spring is pushed at 
constant velocity,
and both the spring horizontal displacement $d(t)$, and the vertical
position of the plate $h(t)$ are monitored during the motion as shown
in Fig.~\ref{illustration}.
The horizontal position $x(t)$ of the plate in the laboratory frame
is deduced from $d(t)$ through the relation $x(t) = V t - d(t)$
(Fig.~\ref{x_vs_t}).
After a transient regime (which is discussed in Sec. III.C),
both the frictional force and
the vertical position of the plate are found to reach steady asymptotic 
values, provided that the spring is pushed at a velocity larger than about  
$1~\mu \rm m/s$.
We focus next on the behavior in this steady-state regime.

\subsection{Steady-state regime}

\begin{figure}
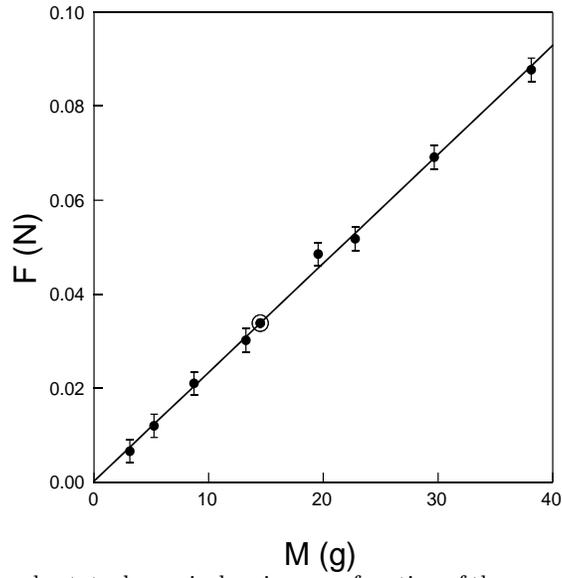

\caption{Frictional force $F$ in the steady state dynamical regime as a 
function of the mass $M$: The slope gives $\mu_d = 0.236\pm0.004$
($k = 189.5~\rm N/m$, $V = 28.17~ \mu \rm m/s$). The circle
denotes the plate used in most of the experiments.}
\label{mu_vs_m}
\end{figure}

The frictional force $F_d$ in the steady-state dynamical regime is
proportional to the mass $M$ of the sliding plate 
as shown in Fig.~\ref{mu_vs_m}.
The frictional force is then written:
\begin{equation}
F_d = \mu_d M g,
\label{frictional_coefficient}
\end{equation}
where $\mu_d$ is the frictional coefficient in the steady-state
regime, which might depend on the plate velocity.
There is no evidence of an additional viscous contribution
to the frictional force, which could arise due to the surrounding water.
Indeed, the linear interpolation leads to
$F_d \simeq 0$ for $M = 0$. [We find $F_d = 0.0002~\rm N$ 
for $M = 0$, a value that is consistent with zero, given the 
uncertainty in $M$.]   It is important to know that such a contribution 
is insignificant,
since it would be independent of the normal applied stress.
The experimental slope in Fig.~\ref{mu_vs_m} leads to a friction 
coefficient $\mu_d = 0.236 \pm 0.004$ for the  $103~\mu \rm m$ diameter beads.
This value is significantly smaller 
than that measured for the same material when dry.  In that case, 
we observe stick-slip motion for which $\mu_d$  is found to vary
between about $0.4$ and $0.6$ \cite{Gollub}.

\begin{figure}
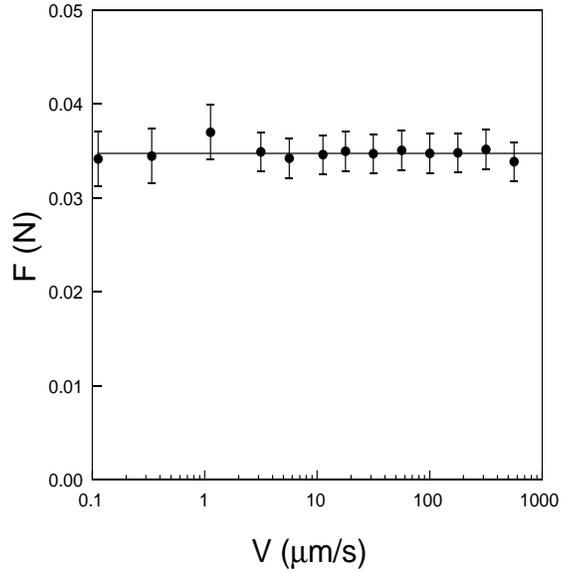

\caption{Frictional force $F$ in the steady-state regime
as a function of the pulling velocity $V$. There is no evidence of
a dependance of the frictional coefficient $\mu_d$ on $V$
($k = 189.5~\rm N/m$, $M = 14.5~\rm g$). }
\label{mu_vs_v}
\end{figure}

We also analyzed the dependence of the frictional force on the
velocity in the steady-state regime that arises after
the initial transient.
Within the experimental resolution, we find that the frictional force does not
depend on the pulling velocity $V$ over four orders of magnitude, 
as shown in Fig.~\ref{mu_vs_v}.



\begin{figure}
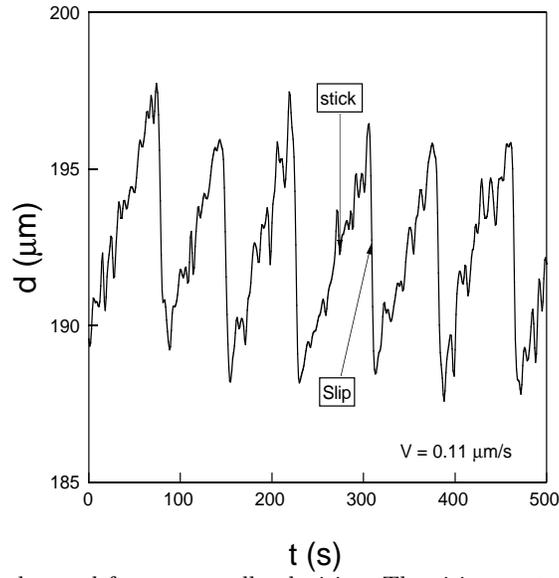

\caption{Typical {\it stick-slip} motion observed for very small velocities.
The rising parts of $d(t)$ correspond to ${dx/dt} = 0$; The plate
is at rest and the applied stress increases linearly with time. The sudden
decreases of $d$ correspond to the slip events; The maximal velocity
is then about $2~\mu \rm m/s$. The amplitude of oscillation is 
about $10~\mu \rm m$ around the mean value $<d> = 192~\mu \rm m$ 
($k = 189.5~\rm N/m$, $M = 14.5~\rm g$, $V = 0.11~\mu \rm m/s$). }
\label{stick-slip}
\end{figure}

However, when the velocity is very small (less than about $0.1 ~\mu \rm m/s$),
we find {\it stick-slip} motion
\cite{explanation} (Fig.~\ref{stick-slip}).
In this case, which corresponds to the leftmost point in Fig.~\ref{mu_vs_v},
we have used the time average frictional force as an estimate. 
The steady state, which could presumably be reached for such small 
velocities by increasing the spring constant, would probably have a 
somewhat higher friction coefficient since velocity weakening is 
a requirement for the stick-slip instability.  
Except for such very low velocities (typically $V < 0.1~\mu \rm m/s$), 
it is safe to conclude that the friction coefficient is independent 
of velocity.  

\subsection{Transient regime}

\subsubsection{Description}

The sequence of events in the transient regime is as follows.  
When the spring first comes into contact with the
rounded tip at {\it A} in Fig.~\ref{illustration}(a), 
the applied stress increases
linearly with time, and the plate
remains at rest (Fig.~\ref{x_vs_t})
until a minimum horizontal applied stress is reached
at {\it B}.
During this initial stress loading over the interval {\it AB},
no significant change in the vertical
position of the plate is observed [Fig.~\ref{illustration}(b)]. 
When the applied stress is large enough, the plate
begins to slide significantly.
The layer begins to dilate while the applied stress 
continues to increase along {\it BC}, reaches a maximum at {\it C}, 
and then decreases along {\it CD} to its asymptotic value at {\it E}.
The total variation of the vertical position of the plate
is only a few micrometers (typically $5~\mu \rm m$).

\subsubsection{Layer dilation}

\begin{figure}
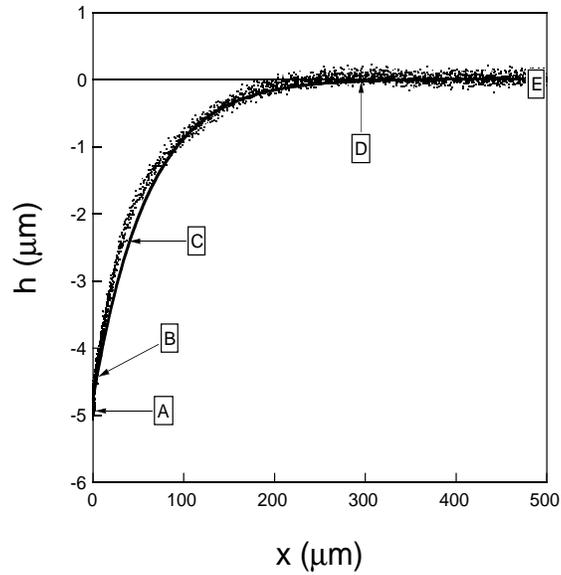

\caption{Vertical displacement $h$ as a function of the 
horizontal displacement $x$ of the plate
($k = 189.5~\rm N/m$, $M = 14.5~\rm g$, $V = 28.17~\mu \rm m/s$).
Dots - experimental points.
Line - exponential interpolation with $R = 59~\mu \rm m$
in Eq.~\ref{dilation_rate}.
The slope is finite at $x=0$.}
\label{h_vs_x}
\end{figure}

\begin{figure}
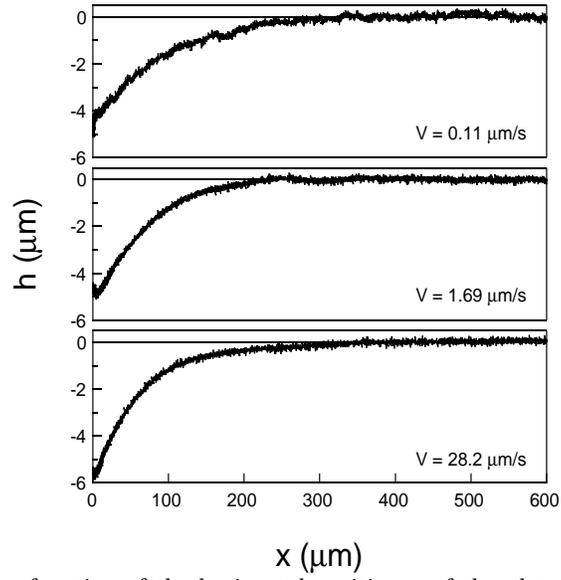

\caption{Vertical position $h(x)$ as a function of the 
horizontal position $x$ of the plate for different velocities.
The granular layer dilates over a distance comparable to the 
bead radius
($k = 189.5~\rm N/m$, $M = 14.5~\rm g$).}
\label{h_vs_v}
\end{figure}

\begin{figure}
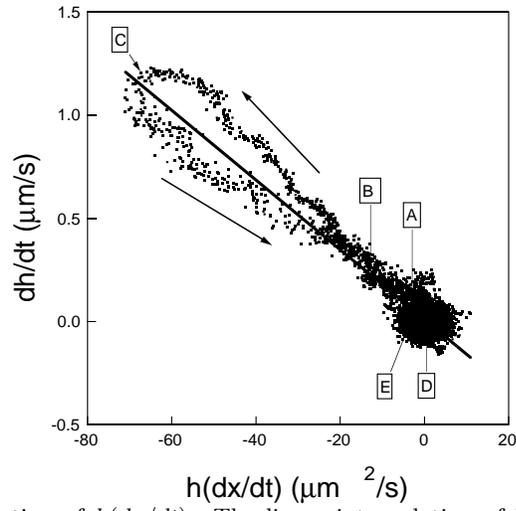

\caption{Dilation rate $dh/dt$ as a function of $h(dx/dt)$.
The linear interpolation of Eq.~\ref{dilation_rate} leads to 
$R \simeq 59~\mu \rm m$, which is comparable to the bead radius
($k = 189.5~\rm N/m$, $M = 14.5~\rm g$, $V = 28.17~\mu \rm m/s$).
The arrows indicate increasing time.}
\label{dilation}
\end{figure}

We find that the vertical position of the
plate tends to its asymptotic value roughly exponentially with the
sliding distance as shown in Fig.~\ref{h_vs_x},
and that the dilation $h(x)$ is roughly independent of the
driving velocity $V$ (Fig.~\ref{h_vs_v}).
This result suggests that the dilation rate $dh/dt$
may be expressed as a function of the dilation $h$ and of the plate velocity
$dx/dt$ as follows:
\begin{equation}
{dh \over dt} = - {h \over R} {dx \over dt},
\label{dilation_rate}
\end{equation}
where $R$ is the characteristic distance over which the layer dilates.
The experimental behavior of $dh/dt$ as a function of $h$ and
$dx/dt$ is shown in Fig.~\ref{dilation}.
Although the law given in Eq.~(\ref{dilation_rate}) is only
roughly satisfied, the mean slope of the curve leads to
$R \simeq 59~\mu \rm m$, which is approximately the mean radius of the 
glass beads.  The corresponding exponential variation of $h(x)$ shown in 
Fig.~\ref{h_vs_x} demonstrates good agreement between
Eq.~(\ref{dilation_rate}) and the experimental data.



We measure no systematic dependence
of the total dilation $\Delta h  = h(\infty) - h(0)$ on the driving
velocity $V$ for layers prepared in the same way. Any variation is, at most,
$1~\mu \rm m$ over the full velocity range accessible to the experimental 
setup.
Since we expect the initial compaction of the layer to
be the reproducible (given identical preparation), we conclude that the 
mean density of the sheared granular
layer does not depend on the shear rate within the experimental resolution.
In contrast, the total dilation $\Delta h$ decreases when the normal 
applied stress
is increased; $\Delta h$ typically decreases
by $0.3~\mu \rm m$ when the mass $M$ of the plate is increased
by $1~\rm g$. 

\begin{figure}
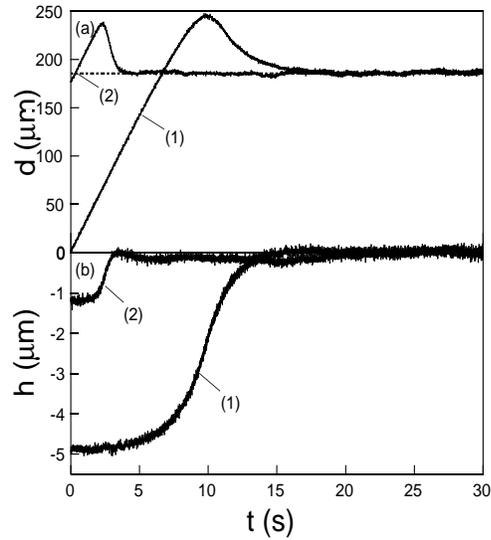

\caption{Behavior (a) of the spring displacement $d(t)$ and
(b) of the vertical position $h(t)$ as functions
of time $t$ in two different cases: (1) The horizontal stress
is released before the experiment;
(2) The horizontal stress is continuously applied.
In the second case, the layer is initially less packed and,
as a consequence, the total dilation $\Delta h$ observed during 
the experiment is less
($k = 189.5~\rm N/m$, $M = 14.5~\rm g$, $V = 28.17~\mu \rm m/s$).}
\label{example}
\end{figure}

The total dilation $\Delta h$ depends strongly on the initial conditions.
For instance, we show in Fig.~\ref{example} the different behaviors
of $h$ and $d$ as functions of time in two cases:
(1) The horizontal stress is released prior to the experiment.
[The layer is prepared as described in Sec. III.A
with results shown in Fig.~\ref{illustration}.]  In this case
we find $\Delta h \simeq 5~\mu \rm m$.
(2) The horizontal stress is not released between runs.
[The plate is initially pushed at constant velocity
($20~\mu \rm m/s$) until the steady-state regime is reached. 
The motion of the translator is suddenly stopped and
the plate stops at a well-defined horizontal applied stress
($F = 3.2 \times 10^{-2}~{\rm N} \simeq F_d$).
The translator motion is then started again.]
In this second case we find that the total dilation of the layer during 
the motion is only 
$\Delta h \simeq 1~\mu \rm m$.
The smaller dilation observed in case (2) suggests that the 
continuously applied horizontal stress prevents the layer from compacting
freely between runs. 

The small magnitude of the total dilation, 
roughly $10\%$ of the bead radius, implies
that the shear zone is localized to at most a few layers of beads. 
The near constancy of the vertical displacement during sliding  suggests 
that shear may organize the beads into horizontal layers. 


In summary, a significant dilation of the granular layer accompanies
the horizontal motion of the plate. During the transient regime,
the layer dilates from its initial compaction to allow 
the horizontal motion of the plate. This slight expansion of less than
one tenth of the bead diameter 
occurs over a characteristic horizontal displacement comparable 
to the bead radius. The total dynamical dilation 
decreases when the normal stress
is increased and does not depend on the driving
velocity. Experiments performed with $200$ and $500~\mu \rm m$ diameter
beads show that the total dilation scales with the bead size.


\subsubsection{Frictional force}

The frictional force $F$ reaches a maximum $F_{max}$
during the initial transient while the layer dilates
(label  {\it C}  in Fig.~\ref{illustration}).
For a given layer under the same experimental
conditions, the measured value of $F_{max}$ is reproducible to within 10\%.
As explained in the following, this experimental scatter
originates essentially in fluctuations of the initial compaction. This 
scatter is small enough not to interfere with a measurement of the maximum
frictional force as a function of $V$ and $\Delta h$.

\begin{figure}
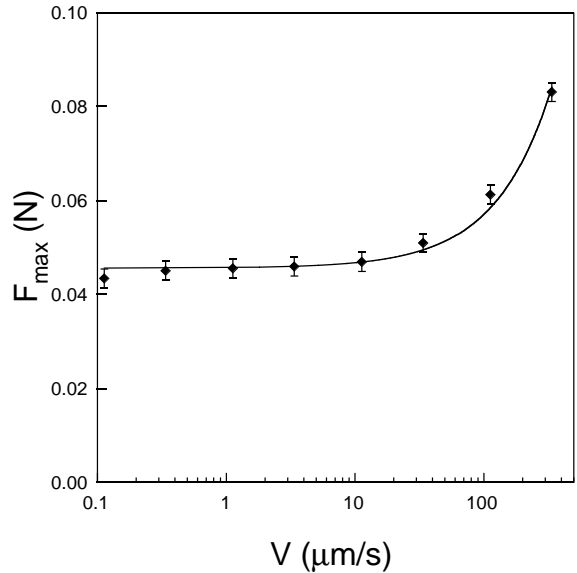

\caption{Maximum frictional force $F_{max}$ as a function of
the driving velocity $V$
($k = 189.5~\rm N/m$, $M = 14.5~\rm g$).}
\label{fmax_vs_v}
\end{figure}

In contrast to the behavior of the steady-state frictional force
$F_d$, the maximum frictional force $F_{max}$ increases with $V$
over the whole range of accessible driving velocities.
Nevertheless, we find that $F_{max}$ increases only slowly
for $V < 100~\mu \rm m/s$ as shown in Fig.~\ref{fmax_vs_v}.

\begin{figure}
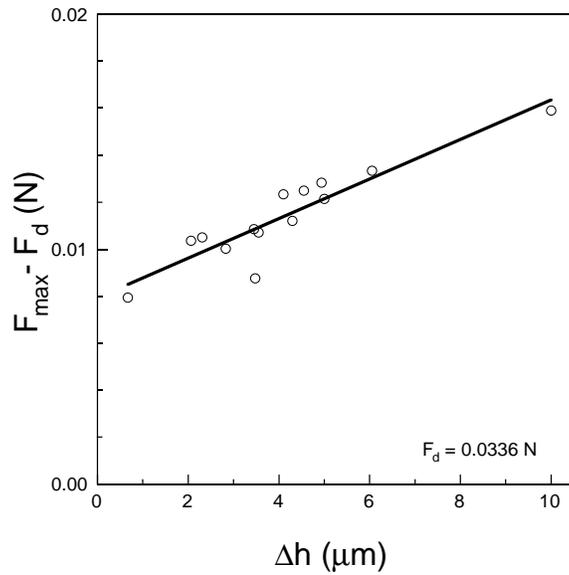

\caption{Maximum frictional force $F_{max}$ as a function 
of the total dilation $\Delta h = h(\infty) - h(0)$.
The layer is initially unstressed
($k = 189.5~\rm N/m$, $M = 14.5~\rm g$, $V = 28.17~\mu \rm m/s$).}
\label{Fmax_vs_h0}
\end{figure}

When layers of new material are prepared, the initial compaction 
fluctuates somewhat.  We infer this fact from the dependence of
the maximum frictional force $F_{max}$ on the total dilation $\Delta h$
for many layers that have not been subjected to an initial horizontal
stress.  The results are plotted in Fig.~\ref{Fmax_vs_h0}, where one can 
see that
the maximum frictional force increases linearly with $\Delta h$:
\begin{equation}
F_{max} = F_d + \alpha + \beta\Delta h,
\label{fm_vs_h}
\end{equation}
whith $\alpha = (7.9 \pm 0.4)~10^{-3}~\rm N$
and $\beta = (8.4 \pm 1.0)~10^{-4}~\rm N/\mu\rm m$.



\begin{figure}
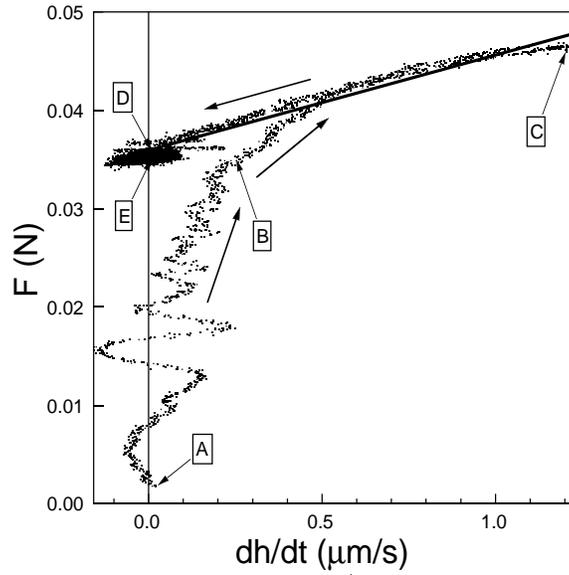

\caption{Frictional force $F$ as a function of the dilation rate
$dh/dt$ showing that $F$ increases roughly linearly
with $dh/dt$ between B and D 
($k = 189.5~\rm N/m$, $M = 14.5~\rm g$, $V = 28.17~\mu \rm m/s$).
The initial oscillations from A  to B  are experimental artifacts due 
to the differentiation of
experimental data containing noise.}
\label{d_vs_dhdt}
\end{figure}

Thus, the initial overshoot of the frictional force is at least
partially related to the additional energy the system requires
to dilate.  
In Fig.~\ref{d_vs_dhdt}, we show the variation 
of the frictional force $F$ with the dilation rate
$dh/dt$ during a single transient event.  As soon as the plate 
moves significantly in the
horizontal plane ({\it B} and afterwards), $F$ increases
roughly linearly with the dilation rate $dh/dt$ and can be described by
\begin{equation}
F \simeq F_d + \nu {dh \over dt},
\label{force}
\end{equation}
with $\nu = (10.7 \pm 0.5)~10^3~\rm Kg/s$.
Note that $F$ depends linearly on the dilation rate $dh/dt$ 
rather than on the dilation $h$ itself. This result is surprising.  
If we suppose that the overshoot of the frictional force originates 
in the potential energy used to lift the plate, one would expect the 
frictional force to depend linearly on $h$: The energy balance between 
the additional potential energy acquired by the weight per unit time 
$U = M g (dh/dt)$ and the power provided by the driving system 
$P = F (dx/dt)$ would lead to 
$F = F_d + M g (dh/dt)/(dx/dt) = F_d - M g h / R$ 
according to Eq.~(\ref{dilation_rate}). 
The experiment performed by Marone {\it et al.} at large stress agrees 
with this prediction \cite{Marone90}. Nevertheless, such a dependence 
is not observed in our experiments. The excess work done by the driving system
 during the overshoot is experimentally about twice the increase in the 
potential 
energy.

We also notice from Fig.~\ref{d_vs_dhdt} that Eq.~(\ref{force}) fails
to describe the initial stage of the motion, when the velocity and
the displacement of the plate remain small (typically 
${dx/dt} < 3~\mu \rm m/s$ and $x < 3~\mu \rm m$). 
The total displacement of the plate is then only $6\%$ of the
bead radius and the velocity is so small that we can expect the
system to respond as an elastic medium. However, the experimental
setup does not currently allow us to study in detail the
very early response of the granular layer to a stress loading and unloading
because of play in the driving system. 


In summary, the experimental results exhibit the important role
played by the dilation of the layer on the frictional force $F$,
which depends roughly linearly on the dilation rate $dh/dt$ 
when the plate experiences a significant motion in the horizontal
plane. As a consequence, the maximum value reached by the frictional
force during the transient for a given driving velocity $V$ depends
linearly on the
total dilation $\Delta h$ of the layer between its initial state
and the steady-state regime. 

\subsection{Response to a static shear stress larger
than the critical value}

Let us now consider the response of a compact granular layer to a 
static horizontal applied force $F$. The layer is prepared
by driving the system in the steady-state regime at large velocity
(typically $20~\mu \rm m/s$), stopping the motor suddenly, and pulling
the spring back until the applied stress is fully released ($F = 0$).
Afterwards, the spring is again pushed ahead but at lower velocity
(typically $1~\mu \rm m/s$) and is stopped at a given value of the applied
stress $F$. 

\begin{figure}
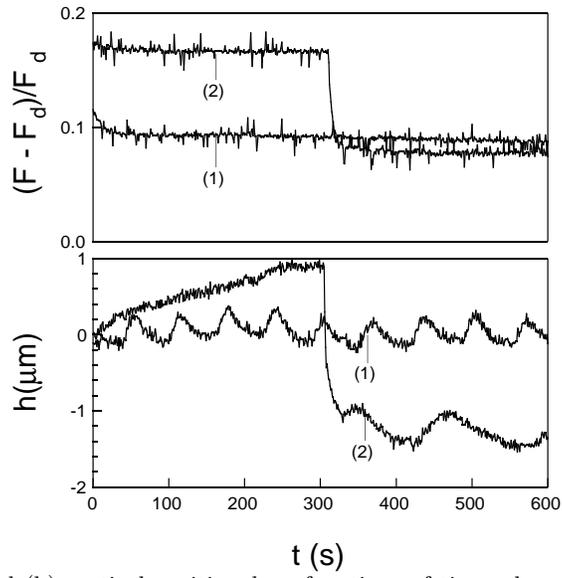

\caption{(a) Applied stress $F$ and (b) vertical position $h$ as functions
of time when a static stress larger than the critical value is applied
($k = 189.5~\rm N/m$, $M = 14.5~\rm g$, $V = 0$).  Two cases are shown:
(1) $F\simeq 1.1~F_d$; (2) $F \simeq 1.17~F_d$.}
\label{statics}
\end{figure}

We find that the granular layer can sustain
the applied stress for a long time (several minutes) when
$F < 1.15~F_d$ typically. Nevertheless, the plate creeps slowly and
the frictional force gradually declines.
Small amplitude oscillations of the vertical position
of the plate are observed (typically $0.2~\mu \rm m$ in amplitude as shown in 
Fig.~\ref{statics}, case 1) without a mean rise of the plate; the
horizontal velocity $dx/dt$ (not shown) does not exhibit measurable
oscillations.  In contrast, when the applied stress is larger 
(e.g. $F \simeq 1.17~F_d$,
case 2 of Fig.~\ref{statics}), the granular layer sustains the stress only 
for a few minutes.
The plate creeps slowly at first
while a significant dilation of the layer is observed 
(typically $1~\mu \rm m$).
After a few minutes a large slip event occurs during
which the stress is released. The plate slides rapidly for a few
tens of micrometers in the horizontal plane and compacts by  
about $2~\mu \rm m$ vertically. 
After the slip event, the horizontal motion of the plate is again
hardly noticeable and vertical oscillations are observed as for lower applied 
stress. The plate can then sustain the remaining stress, which is now smaller,
for a long time.
We do not understand the oscillations of the vertical position of the
plate, but they cannot be experimental
artifacts, to the best of our knowledge.

\section{DISCUSSION AND CONCLUSION}
\subsection{Recapitulation of the main experimental results
and empirical model}

The experiments performed on immersed granular material
allow us to propose an empirical model for the mechanical behavior
of a sheared granular layer under very low stress. The model may also be 
applicable
to dry materials but a stiffer measurement system would be required to 
verify this.  The main results are as follows:

\subsubsection{Layer dilation}

Any horizontal motion of the plate involves a dilation of the
granular layer:

(i) The vertical position of the plate $h$ tends roughly
exponentially (Fig.~\ref{h_vs_x}) to its asymptotic value
over a distance $R$, which is approximately the bead radius.
The dilation rate $dh/dt$ then obeys Eq.~(\ref{dilation_rate}).

(ii) The total dilation $\Delta h$ of the layer does not
depend on the driving velocity $V$ when the plate slides
continuously. 

(iii) The total dilation $\Delta h$ of the layer decreases when
the normal applied stress is increased.

(iv) The total dilation $\Delta h$ of the layer scales like the bead size.

\subsubsection{Frictional force}

After a transient, the plate generally slides continuously.
The measured frictional force in the steady-state regime
 is proportional to the normal applied stress (Fig.~\ref{mu_vs_m})
and does not depend on the driving velocity $V$ over the
whole velocity range accessible to the experimental setup
(Fig.~\ref{mu_vs_v}). Nevertheless, the existence of a
stick-slip motion (Fig.~\ref{stick-slip}) and the response
of the granular layer to a static applied stress (Fig.~\ref{statics})
are consistent with a velocity weakening of the granular layer
at very small velocity (typically $0.1~ \mu m/s$).

During the transient, while the layer is dilating significantly, 
the frictional force $F$ depends roughly linearly on the
dilation rate $dh/dt$ (Eq.~(\ref{force}) and Fig.~\ref{d_vs_dhdt})
and the frictional force reaches a maximum value that increases
with the driving velocity $V$ (Fig.~\ref{fmax_vs_v}) and increases
linearly with the total dilation $\Delta h$ of the layer
(Eq.~(\ref{fm_vs_h}) and Fig.~\ref{Fmax_vs_h0}).

\subsubsection{Empirical model}

The motion of the plate is approximately governed by the differential
equation given by Eq.~(\ref{diff}) in which $F$ must be replaced
by its empirical expression proposed in Eq.~\ref{force}.
The vertical position of the plate obeys Eq.~\ref{dilation_rate}.
The system of differential equations that governs the time evolution
of $x$ and $h$ can be written:

\begin{eqnarray}
M {\frac{d^2 x}{dt^2}}& =& k (V t - x) - F_d - \nu {\frac{dh}{dt}}
\nonumber \\
{\frac{dh}{dt}} &=& - {h \over R} {\frac{dx}{dt}}
\label{system}
\end{eqnarray}
with the initial conditions:
\begin{eqnarray}
x(0) = 0&;&{\frac{dx}{dt}}(0) = 0 \nonumber \\
h(0) &=& -h_0. 
\label{ic}
\end{eqnarray}
The initial value $-h_0$ of $h$ is provided by the experiment.
The system of differential equations [Eqs.~(\ref{system}) and (\ref{ic})]
is then integrated numerically using the Runge-Kutta method
\cite{num_rec}.

In the next section, we discuss the results of our simplified empirical 
model and compare them to the experimental measurements.

\subsubsection{Comparison of the empirical model to the experimental results}

We show in Fig.~\ref{theory} the result of integrating
the empirical model and comparing it to the experimental
data. The model is expected to describe the dynamics of the plate 
only from  {\it B} to  {\it E}. We take as the initial 
value of $h$ its value at {\it B}.
The model correctly describes the dilation of the layer. 
However, the agreement between the experimental and
the theoretical instantaneous values of the force is imperfect;
the small discrepancy is mainly due to the fact that the
model Equation~(\ref{dilation_rate}) does not account for the detailed
behavior of $dh/dt$ in the cycle shown in Fig.~(\ref{dilation}).
Nevertheless, the empirical model is in qualitative agreement
with the experimental results and allows one to predict an increase
of the maximum frictional force $F_{max}$ with the total
dilation $\Delta h$ and the driving velocity $V$.


On the other hand, the empirical model cannot describe 
the very early stage {\it A} to {\it B}
of the plate motion and the response of the granular layer
to a static applied stress. Further experimental studies would 
be required to obtain a complete description of these phenomena.
Our experimental setup does not
allow us to perform these experiments in its present configuration.

\subsection{Comparison to previous work}

Both this work and previous studies \cite{Marone} point out the important 
role played by the dilation on the frictional properties of the sheared 
granular material.  However, the results obtained at small stress differ 
significantly from previous experimental results obtained at much larger 
stress. 

To explain these differences, we first summarize the friction law given by 
Marone {\it et al.} \cite{Marone90}. The first law decribes the dependence 
of the frictional force on the slip velocity $V$ and on the surface's slip 
history {\it via} a state variable $\Psi$ \cite{Rice}:
\begin{equation}
F = M g \left( \mu_0 + b \Psi + a \ln{V \over V^*} \right),
\label{eq1}
\end{equation}
where $F$ is the shear stress, $M g$ is the normal applied stress, $\mu_0$ is 
a constant (which can be understood as the overall frictional coefficient), 
$V^*$ is an arbitrary reference velocity, and $a$ and $b$ are two empirical 
constants. The evolution of the state variable $\Psi$ is governed by
\begin{equation}
{d\Psi \over dt} = -{V \over R} \left( \Psi + \ln {V \over V^*} \right),
\label{eq2}
\end{equation}
where $R$ is a characteristic distance over which the frictional force 
changes following a change in the slip velocity $V$. According to 
Eqs.~(\ref{eq1}) and (\ref{eq2}), the steady-state value of the frictional 
force $F_d$ depends on $V$ and $dF_d/d(\ln V) = a-b$. The experiments 
performed at large normal applied stress qualitatively agree with this 
theoretical description.

We find experimentally that the steady-state value of the frictional 
force $F_d$ does not depend on the driving velocity $V$ at low stress. 
In the absence of any velocity strengthening or weakening, $a = b$ so that 
$dF_d/d(\ln V) = 0$, and the frictional force reads
\begin{equation}
F= F_d - a M g R {d\Psi/dt \over V},
\label{eq3}
\end{equation}
where we have set $M g \mu_0 = F_d$ because $d\Psi/dt = 0$ in the 
steady-state regime.
Eq.~\ref{eq3} agrees with the energetic argument (equality of work done 
and potential energy gain) mentioned in Sec.III.C.3 provided that
$a \Psi = - h/R$. We tentatively assume this connection between the state 
variable $\Psi$ and the vertical displacement $h$ in our work in order to 
allow an interpretation of our results in term of their theory. With this 
assumption the model of Marone {\it et al.} leads to 
$F = F_d + M g {dh \over dx}$. Our experiments performed at small applied 
stress disagree qualitatively with these results. Indeed, the frictional 
coefficient is found to increase linearly with the dilation rate $dh/dt$ 
(Eq.~\ref{force}) rather than with $dh/dx$. However, in making a comparison 
one should also note that the  normal and tangential forces are applied 
independently to the granular material in our work, while in 
Ref.~\cite{Marone90}, the two are equal.

\subsection{Conclusion}

The immersion of the material presents several experimental advantages.
First, it allows one to work in well-defined conditions
and to suppress any variability related to humidity changes.
Efforts to eliminate water are rarely adequate because of 
adsorption. Second, a continuous motion of the sliding plate is
observed even at low driving velocity (that is, stick-slip motion is avoided)
and an extremely precise
study of the steady frictional properties of the granular
layer is then possible. The transient behavior that precedes the steady-state 
continuous motion allows a precise study
of the dynamics of the frictional force for granular materials under low 
stress. A comparison with previous work shows that there are significant 
differences from the high stress case most relevant to geophysics. 

Certainly, there are different physical processes at work at high pressures, 
where the individual particles can be fractured by the stress and plastic 
flow may also occur.  
We hope to obtain additional insight by imaging the granular layer during 
the motion of the plate. The response of the granular layer on very long 
time scales to static applied stress, including the slow strengthening of the 
material in the presence of stress, will  be  analyzed in a further 
publication \cite{Geminard}. We believe that our proposed friction law would 
 also apply to dry material at low stress, although further experiments 
would be required to demonstrate this.

{\bf Acknowledgements}
This work was supported by the National Science Foundation under Grant No.
DMR-9704301. 
J.-C. G. thanks the Centre National de la Recherche Scientifique (France) 
for supporting the research of its members, that was
carried out in foreign laboratories.  
We are grateful for the
collaboration of S. Nasuno, who built the apparatus and developed many of 
the methods used in this investigation. We appreciate helpful
discussions with C. Marone and comments on the manuscript by T. Shinbrot.

\begin{figure}
\caption{Experimental behavior compared with the theoretical model
(a) of the spring displacement $d(t)$ and
(b) of the vertical position $h(t)$ as a function
of time $t$ ($k = 189.5~\rm N/m$, $M = 14.5~\rm g$,
$V = 28.17~ \mu \rm m/s$). dots - experimental data.
lines - empirical model.}
\label{theory}
\end{figure}

\pagestyle{empty} \setcounter{figure}{0}

\pagebreak
\begin{figure}[h]
\centerline{
\epsfig{file=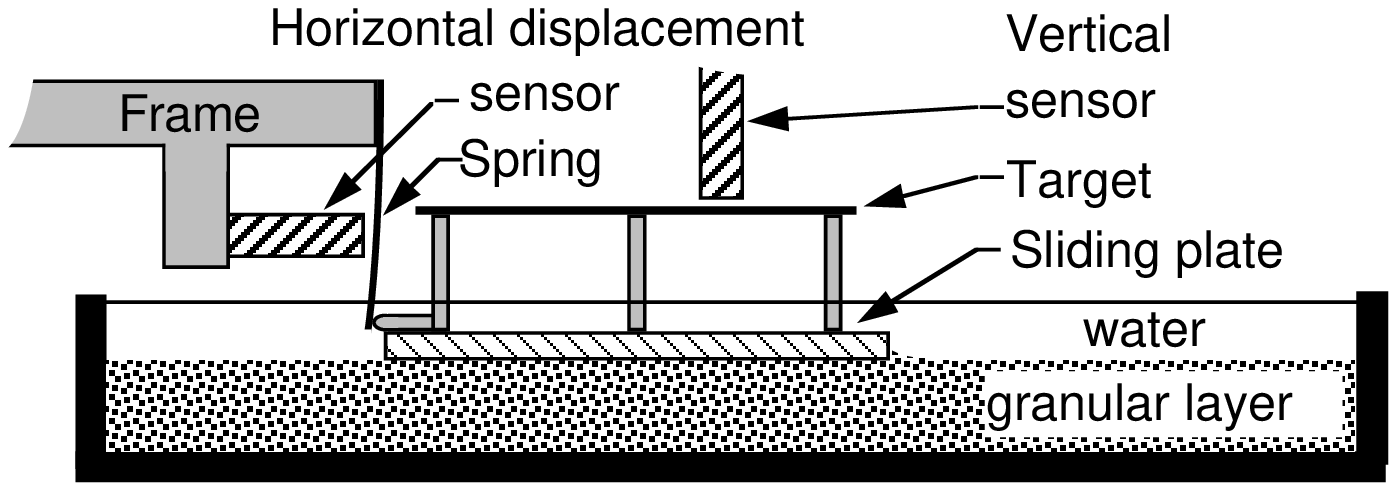,width=18pc,height=9pc,angle=0}
}
\caption{Schematic diagram of the experimental setup for studying a 
fluid-saturated
granular layer with sensitive measurement of horizontal and vertical 
positions.}
\end{figure}

\begin{figure}[h]
\centerline{
\epsfig{file=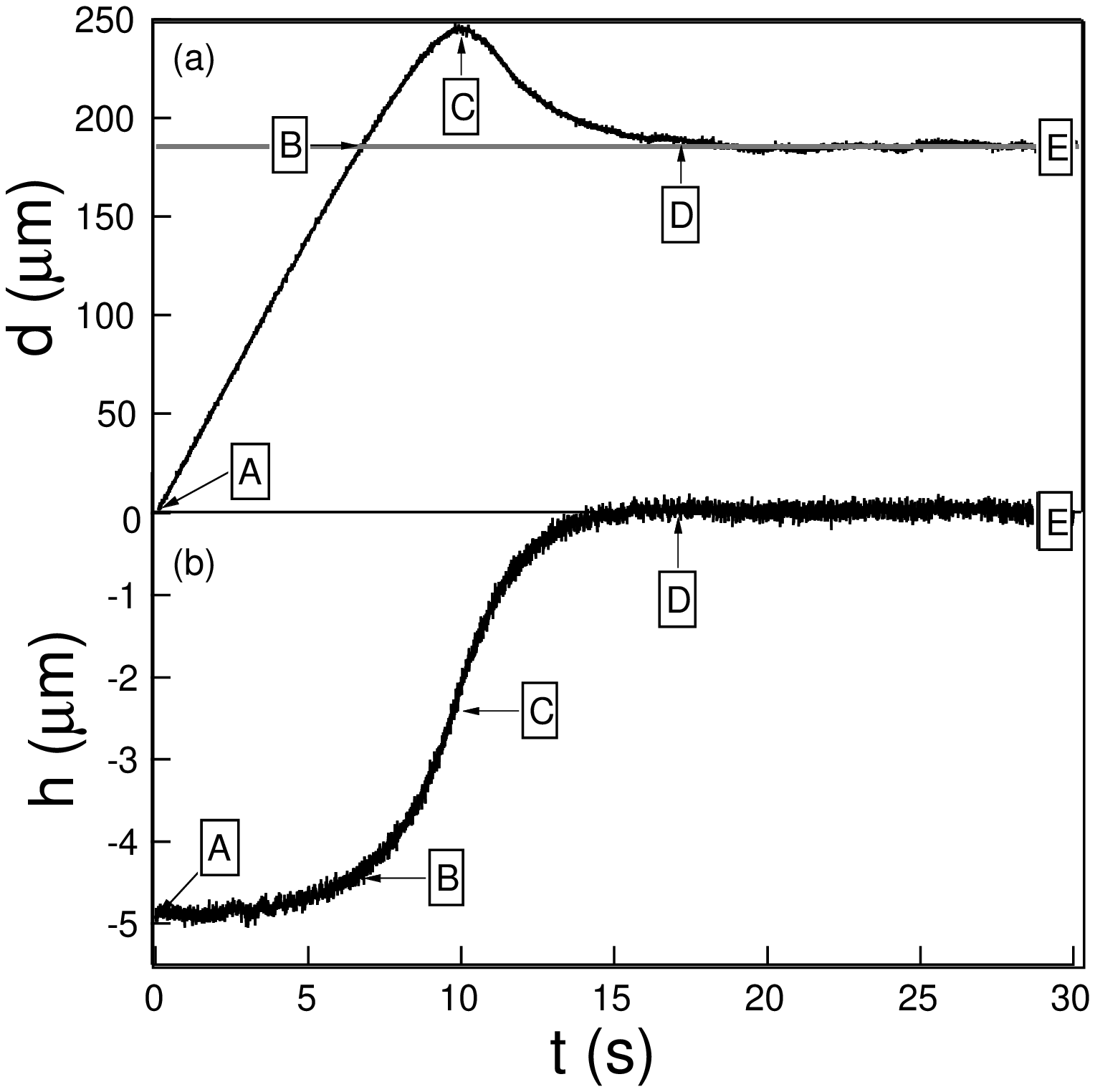,width=18pc,height=18pc,angle=0}
}
\caption{Typical behavior (a) of the spring displacement $d(t)$ and
(b) of the vertical position $h(t)$ as a function
of time $t$ ($k = 189.5~\rm N/m$, $M = 14.5~\rm g$,
$V = 28.17~ \mu \rm m/s$). [The different stages of the plate
motion, labelled A-E, are discussed in section III.C.1.]}
\end{figure}

\begin{figure}[h]
\centerline{
\epsfig{file=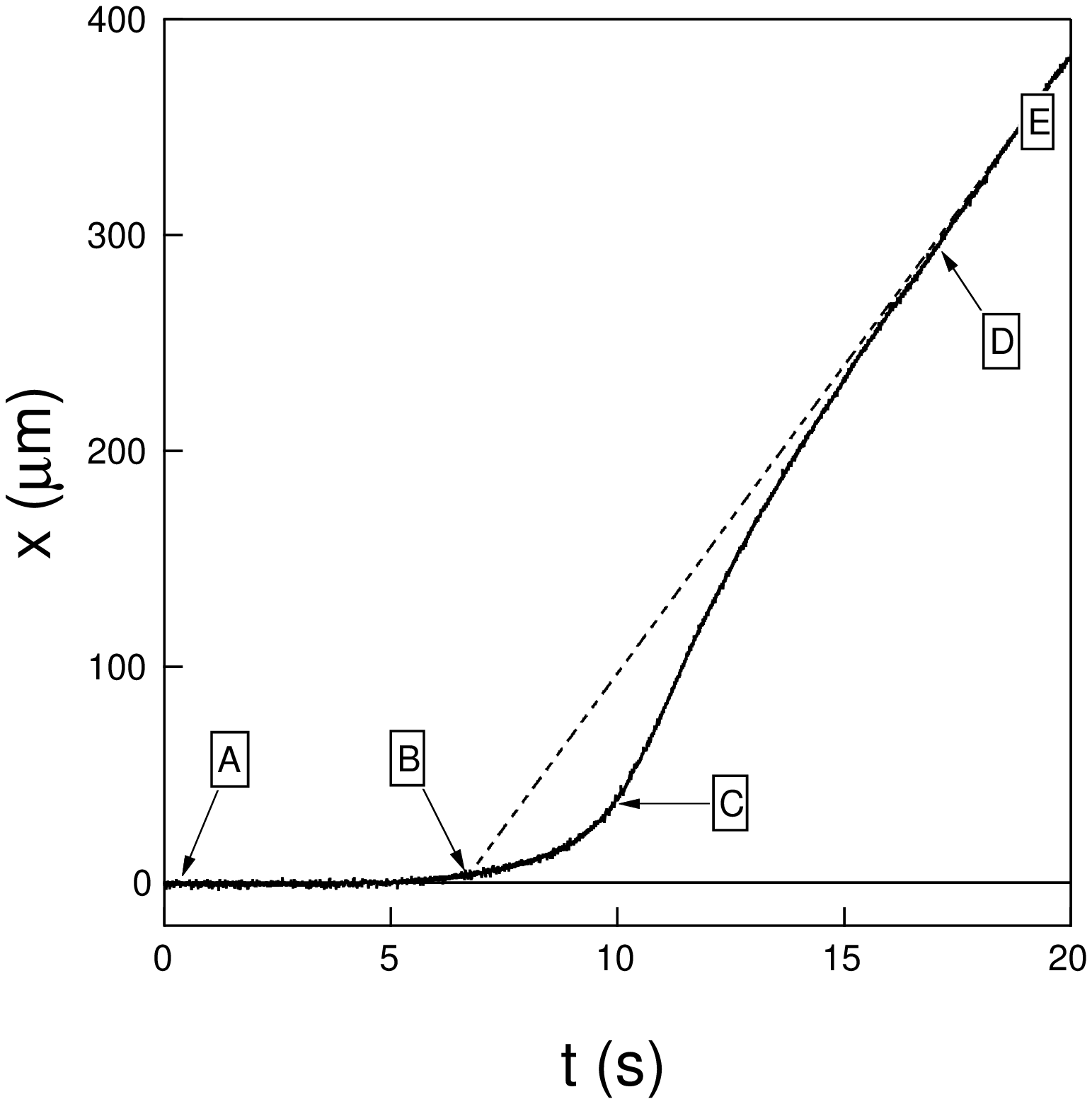,width=18pc,height=18pc,angle=0}
}
\caption{Position of the plate $x$ as a function of time
 $t$ ($k = 189.5~\rm N/m$, $M = 14.5~\rm g$,
$V = 28.17~ \mu \rm m/s$). The dashed line
indicates the asymptotic regime.}
\end{figure}

\begin{figure}[h]
\centerline{
\epsfig{file=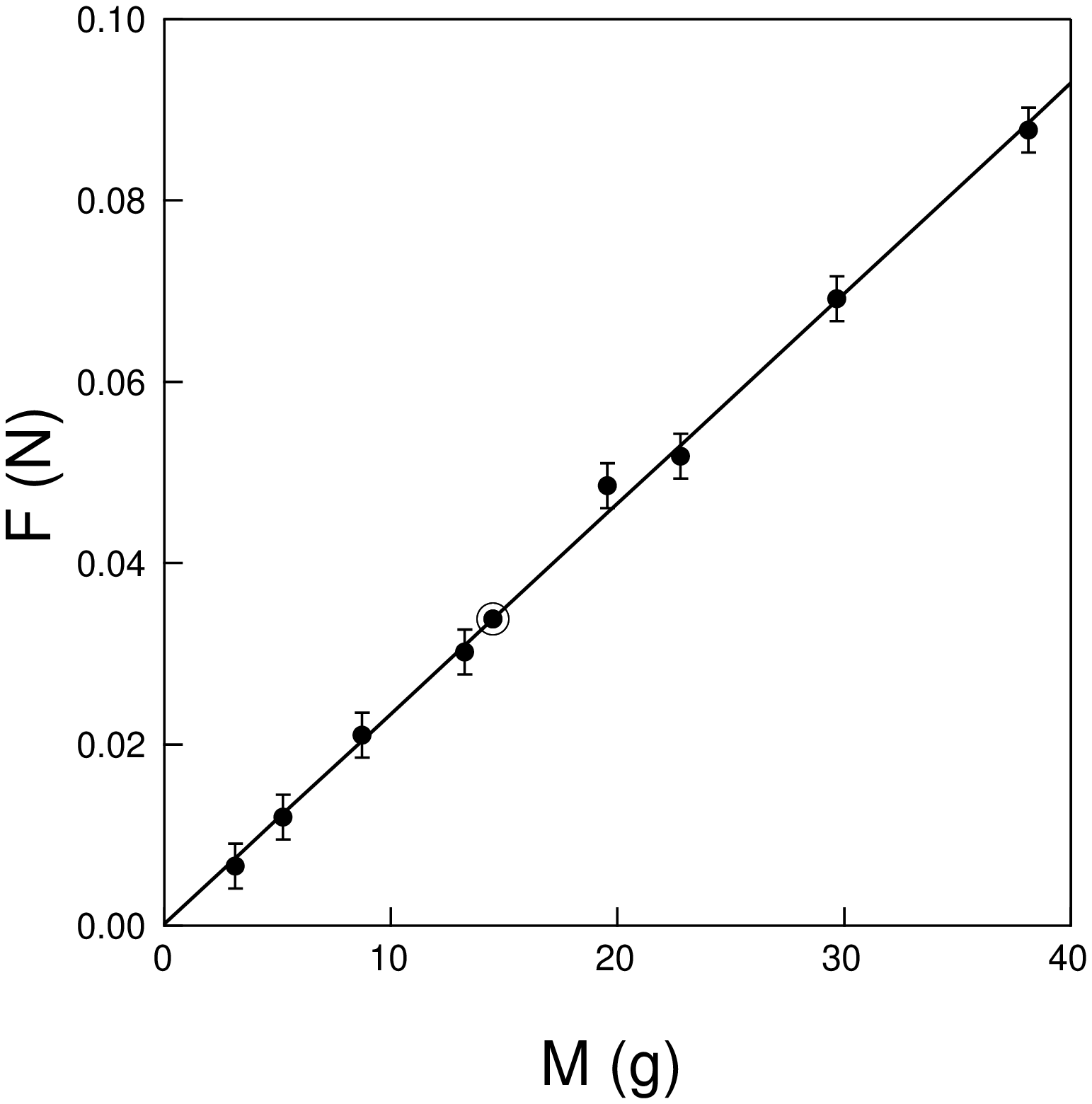,width=18pc,height=18pc,angle=0}
}
\caption{Frictional force $F$ in the steady state dynamical regime as a 
function of the mass $M$: The slope gives $\mu_d = 0.236\pm0.004$
($k = 189.5~\rm N/m$, $V = 28.17~ \mu \rm m/s$). The circle
denotes the plate used in most of the experiments.}
\end{figure}

\begin{figure}[h]
\centerline{
\epsfig{file=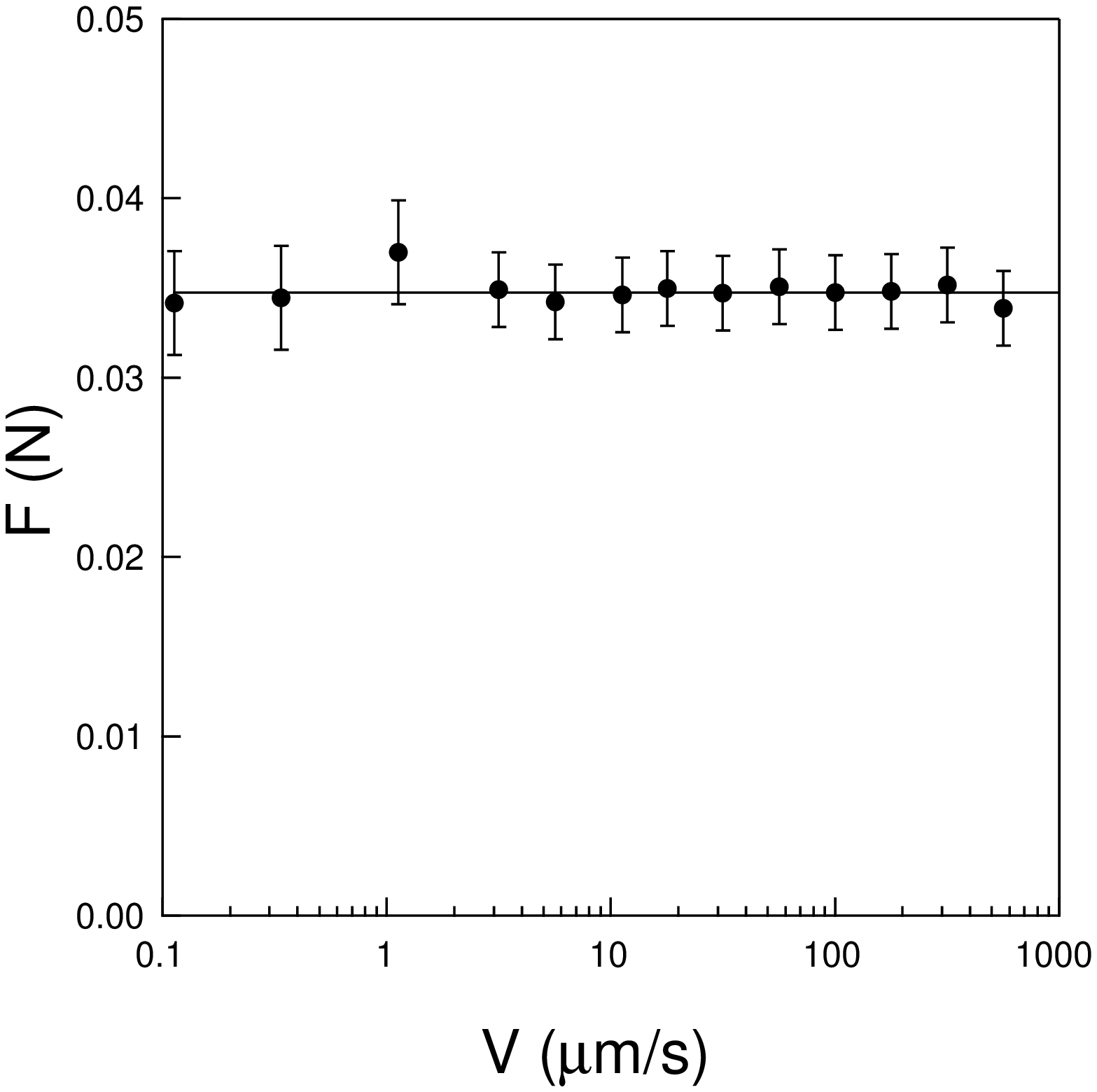,width=18pc,height=18pc,angle=0}
}
\caption{Frictional force $F$ in the steady-state regime
as a function of the pulling velocity $V$. There is no evidence of
a dependance of the frictional coefficient $\mu_d$ on $V$
($k = 189.5~\rm N/m$, $M = 14.5~\rm g$). }
\end{figure}

\begin{figure}[h]
\centerline{
\epsfig{file=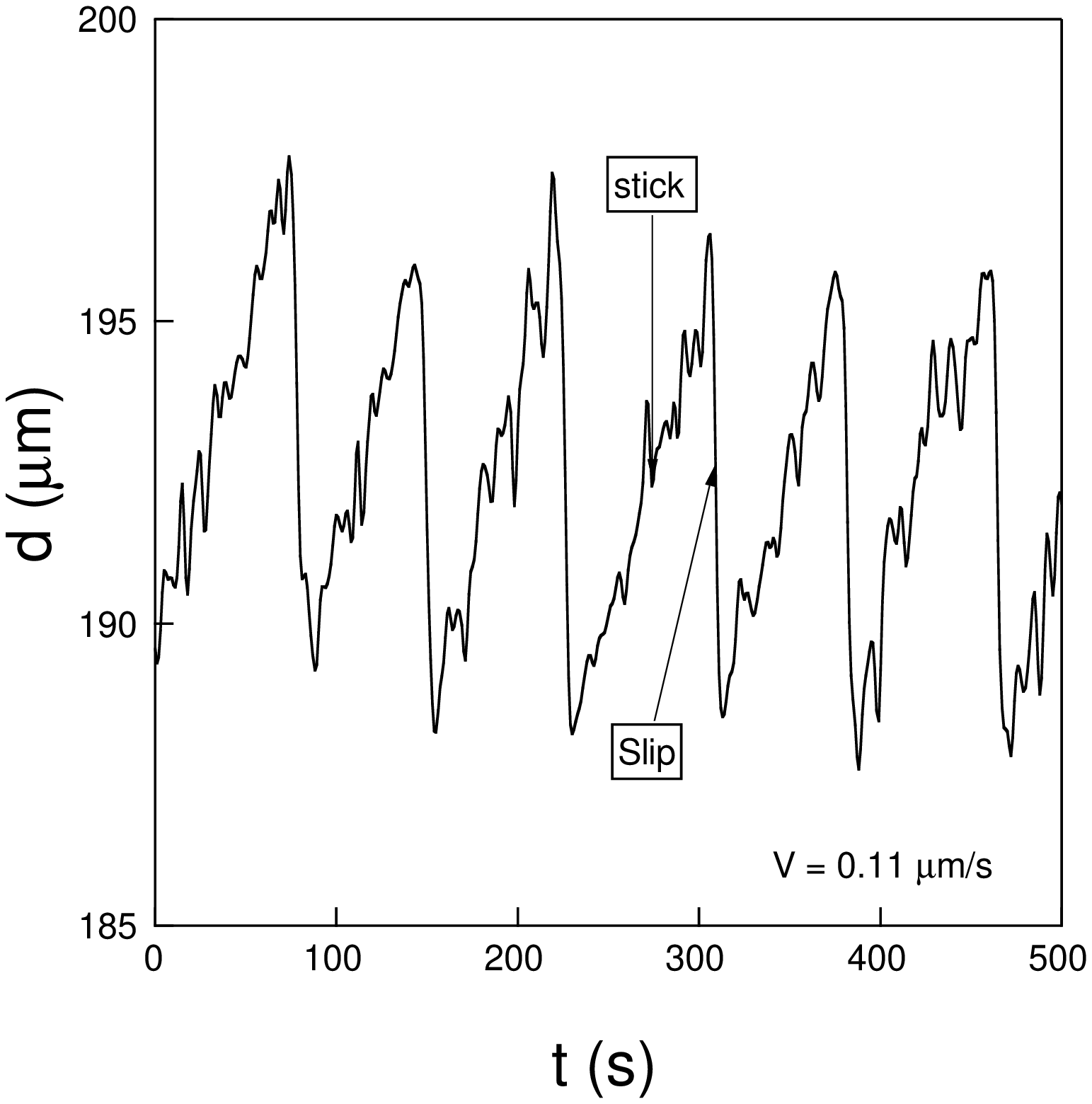,width=18pc,height=18pc,angle=0}
}
\caption{Typical {\it stick-slip} motion observed for very small velocities.
The rising parts of $d(t)$ correspond to ${dx/dt} = 0$; The plate
is at rest and the applied stress increases linearly with time. The sudden
decreases of $d$ correspond to the slip events; The maximal velocity
is then about $2~\mu \rm m/s$. The amplitude of oscillation is 
about $10~\mu \rm m$ around the mean value $<d> = 192~\mu \rm m$ 
($k = 189.5~\rm N/m$, $M = 14.5~\rm g$, $V = 0.11~\mu \rm m/s$). }
\end{figure}

\begin{figure}[h]
\centerline{
\epsfig{file=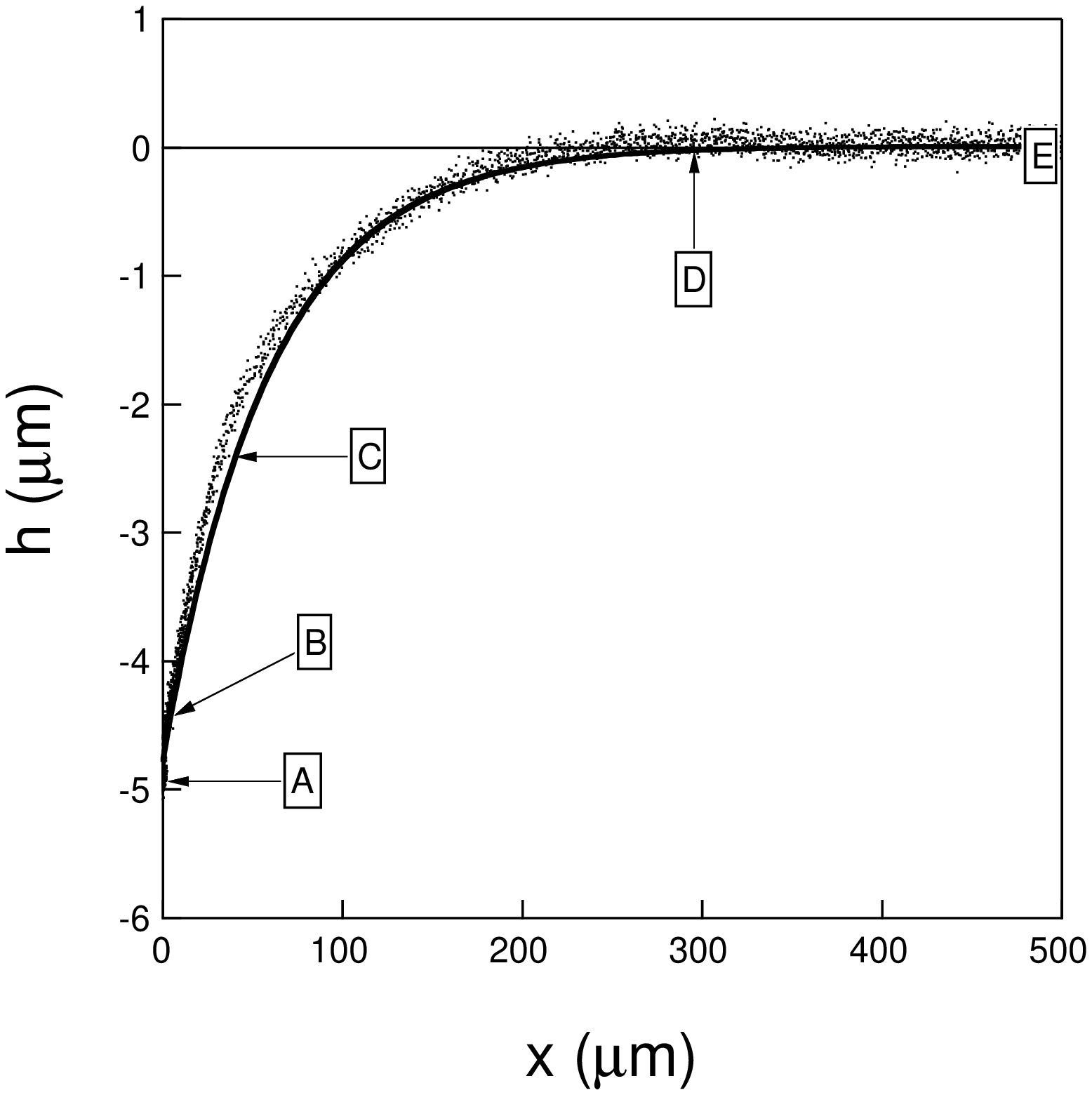,width=18pc,height=18pc,angle=0}
}
\caption{Vertical displacement $h$ as a function of the 
horizontal displacement $x$ of the plate
($k = 189.5~\rm N/m$, $M = 14.5~\rm g$, $V = 28.17~\mu \rm m/s$).
Dots - experimental points.
Line - exponential interpolation with $R = 59~\mu \rm m$
in Eq.~\ref{dilation_rate}.
The slope is finite at $x=0$.}
\end{figure}

\begin{figure}[h]
\centerline{
\epsfig{file=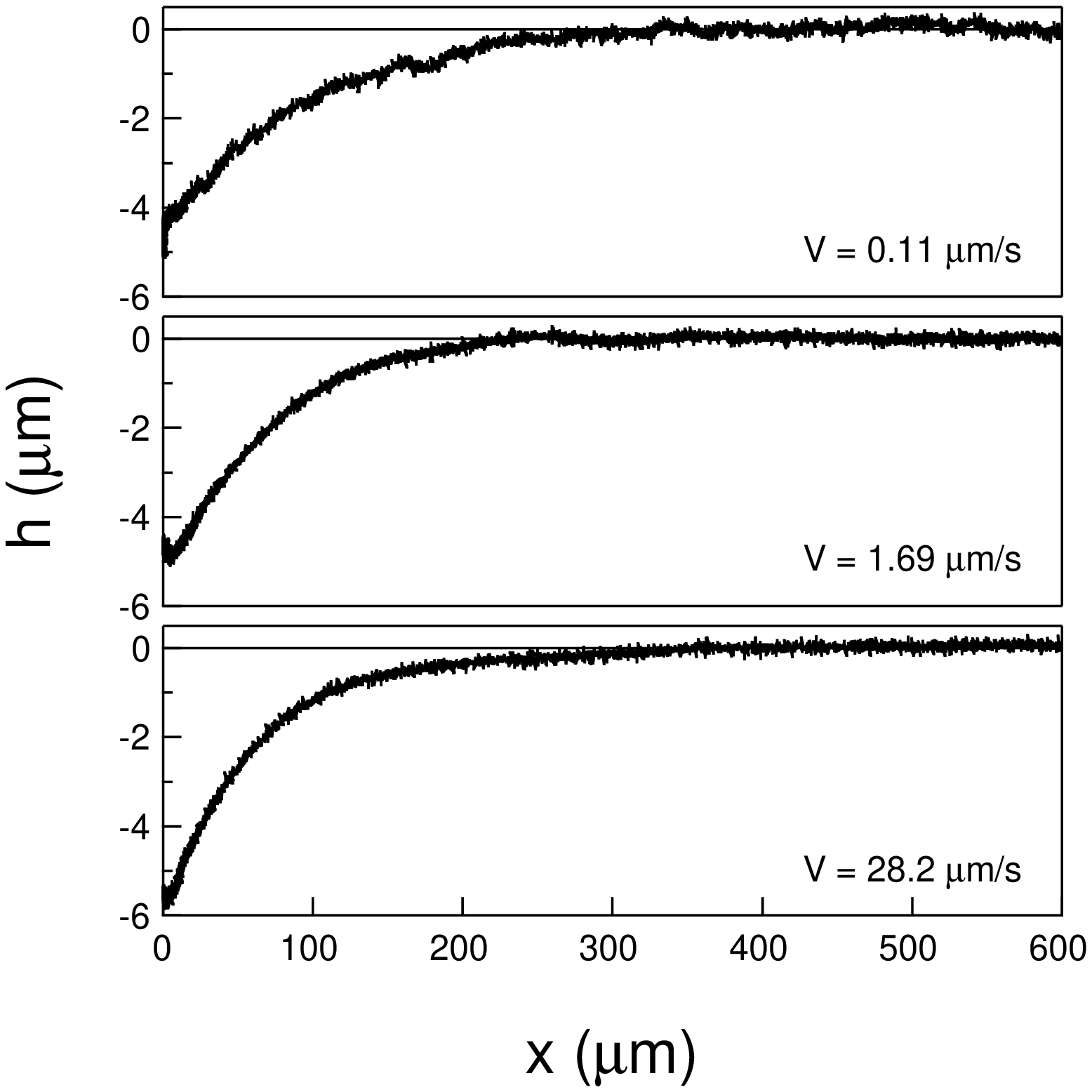,width=18pc,height=18pc,angle=0}
}
\caption{Vertical position $h(x)$ as a function of the 
horizontal position $x$ of the plate for different velocities.
The granular layer dilates over a distance comparable to the 
bead radius
($k = 189.5~\rm N/m$, $M = 14.5~\rm g$).}
\end{figure}

\begin{figure}[h]
\centerline{
\epsfig{file=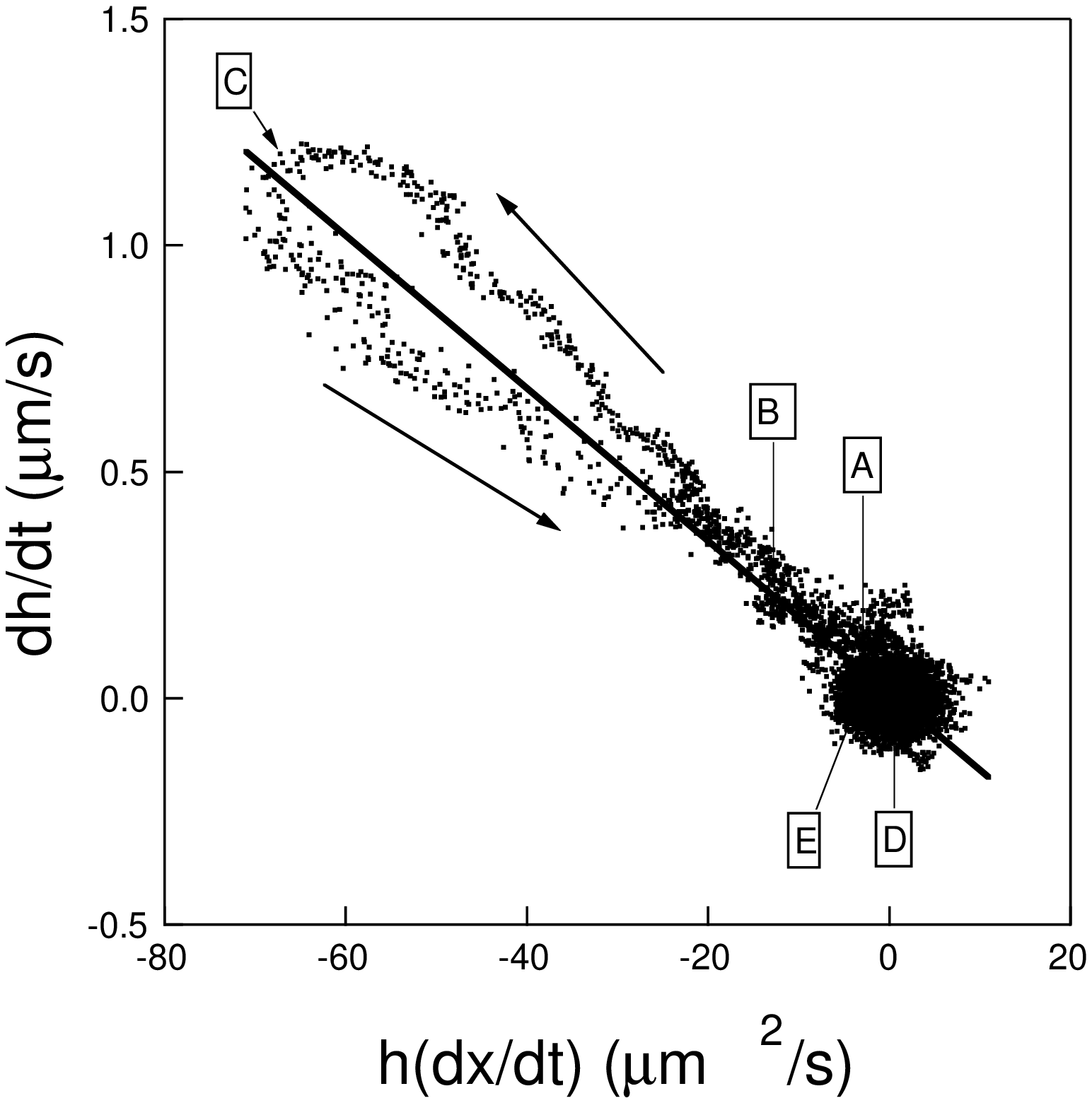,width=18pc,height=18pc,angle=0}
}
\caption{Dilation rate $dh/dt$ as a function of $h(dx/dt)$.
The linear interpolation of Eq.~\ref{dilation_rate} leads to 
$R \simeq 59~\mu \rm m$, which is comparable to the bead radius
($k = 189.5~\rm N/m$, $M = 14.5~\rm g$, $V = 28.17~\mu \rm m/s$).
The arrows indicate increasing time.}
\end{figure}

\begin{figure}[h]
\centerline{
\epsfig{file=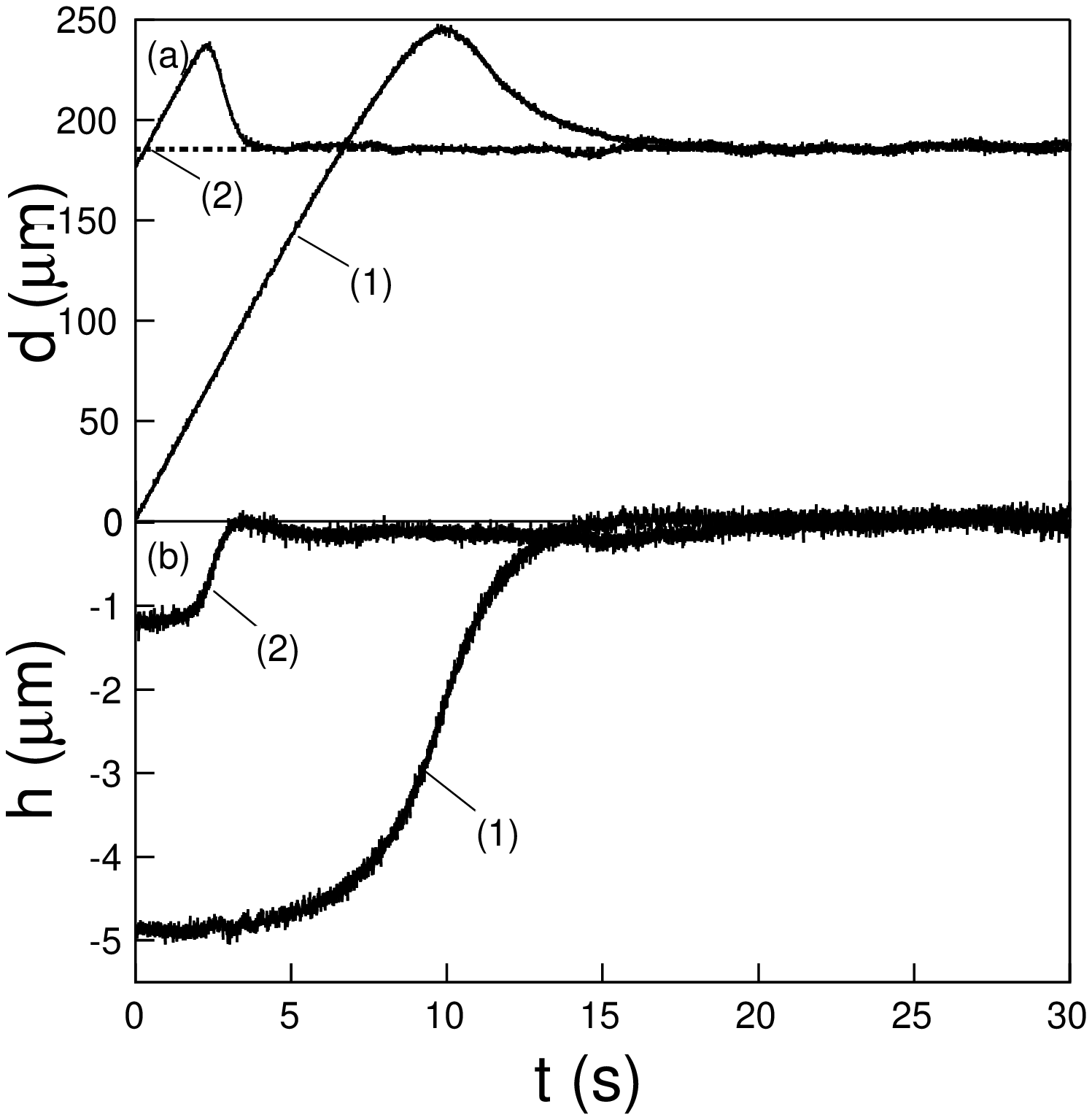,width=18pc,height=18pc,angle=0}
}
\caption{Behavior (a) of the spring displacement $d(t)$ and
(b) of the vertical position $h(t)$ as functions
of time $t$ in two different cases: (1) The horizontal stress
is released before the experiment;
(2) The horizontal stress is continuously applied.
In the second case, the layer is initially less packed and,
as a consequence, the total dilation $\Delta h$ observed during 
the experiment is less
($k = 189.5~\rm N/m$, $M = 14.5~\rm g$, $V = 28.17~\mu \rm m/s$).}
\end{figure}

\begin{figure}[h]
\centerline{
\epsfig{file=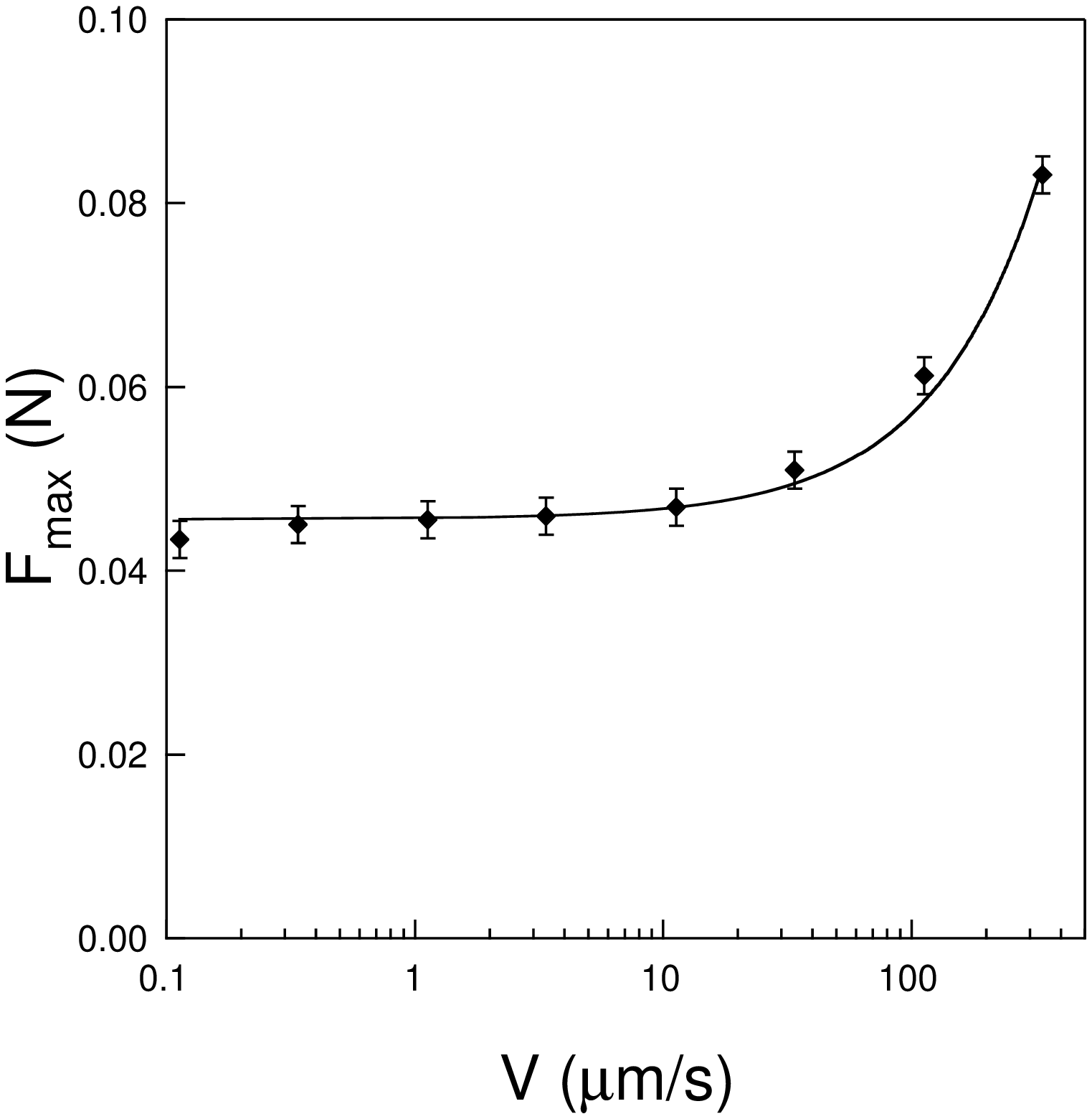,width=18pc,height=18pc,angle=0}
}
\caption{Maximum frictional force $F_{max}$ as a function of
the driving velocity $V$
($k = 189.5~\rm N/m$, $M = 14.5~\rm g$).}
\end{figure}

\begin{figure}[h]
\centerline{
\epsfig{file=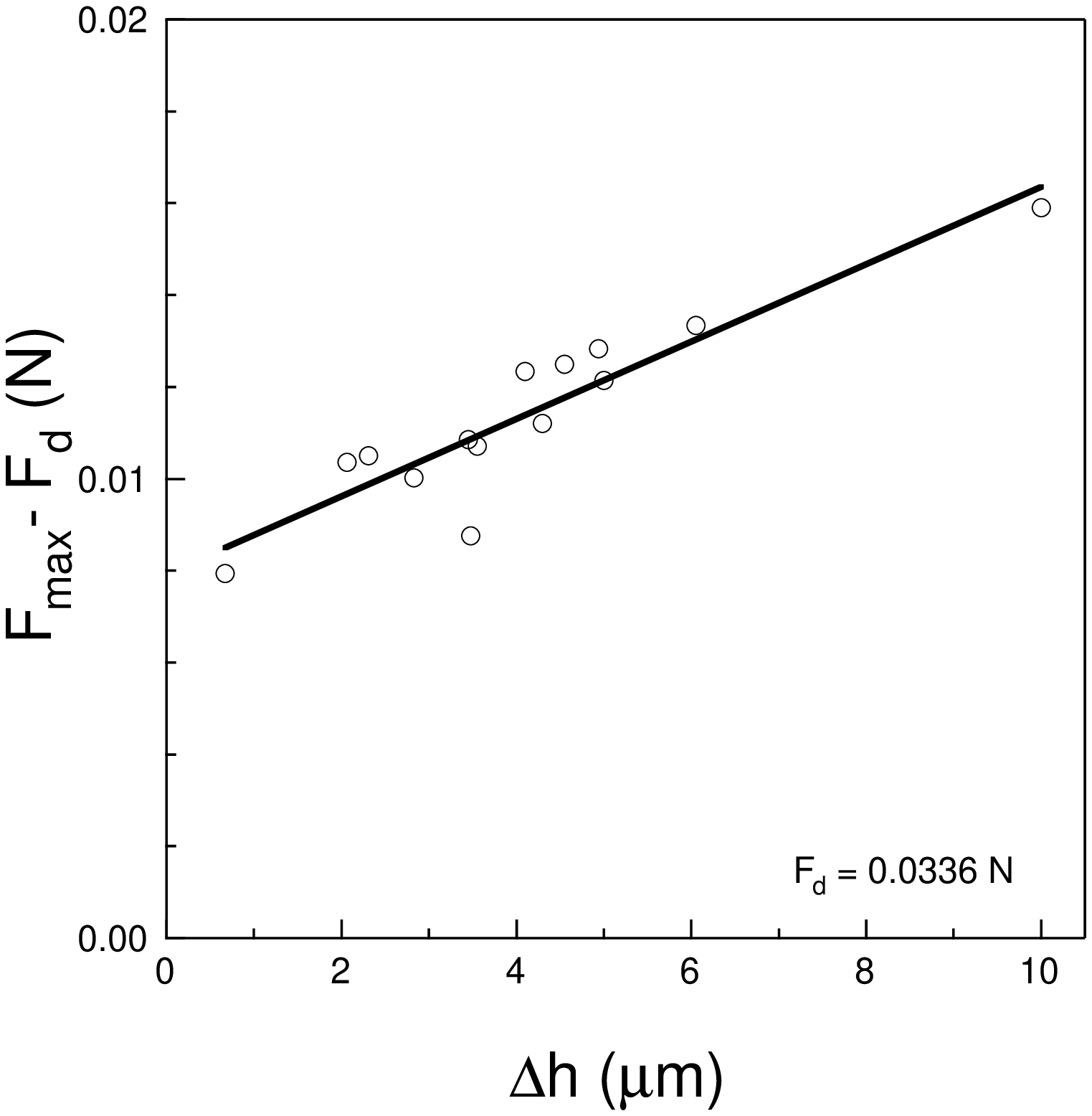,width=18pc,height=18pc,angle=0}
}
\caption{Maximum frictional force $F_{max}$ as a function 
of the total dilation $\Delta h = h(\infty) - h(0)$.
The layer is initially unstressed
($k = 189.5~\rm N/m$, $M = 14.5~\rm g$, $V = 28.17~\mu \rm m/s$).}
\end{figure}

\begin{figure}[h]
\centerline{
\epsfig{file=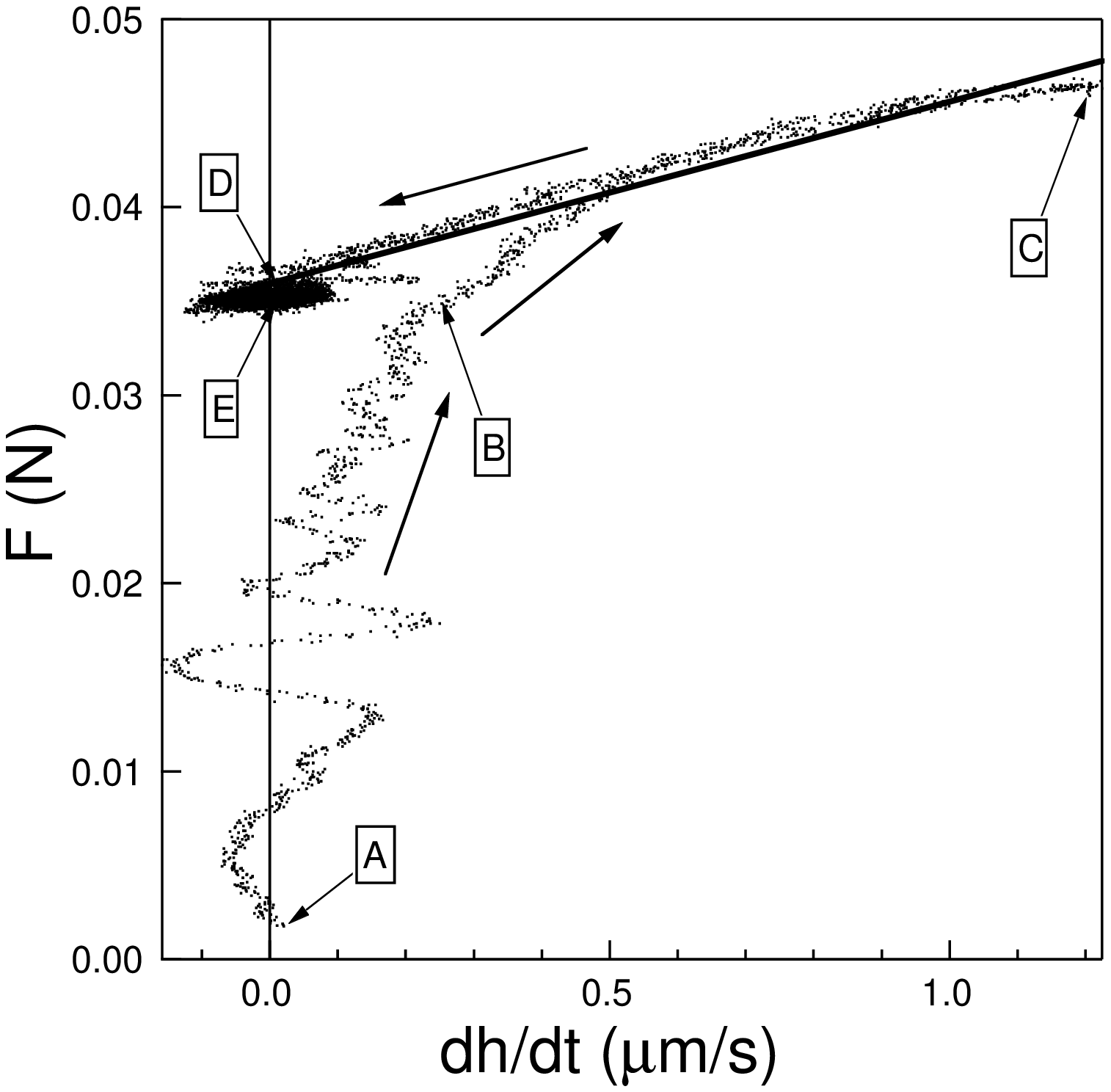,width=18pc,height=18pc,angle=0}
}
\caption{Frictional force $F$ as a function of the dilation rate
$dh/dt$ showing that $F$ increases roughly linearly
with $dh/dt$ between B and D 
($k = 189.5~\rm N/m$, $M = 14.5~\rm g$, $V = 28.17~\mu \rm m/s$).
The initial oscillations from A  to B  are experimental artifacts due 
to the differentiation of
experimental data containing noise.}
\end{figure}

\begin{figure}[h]
\centerline{
\epsfig{file=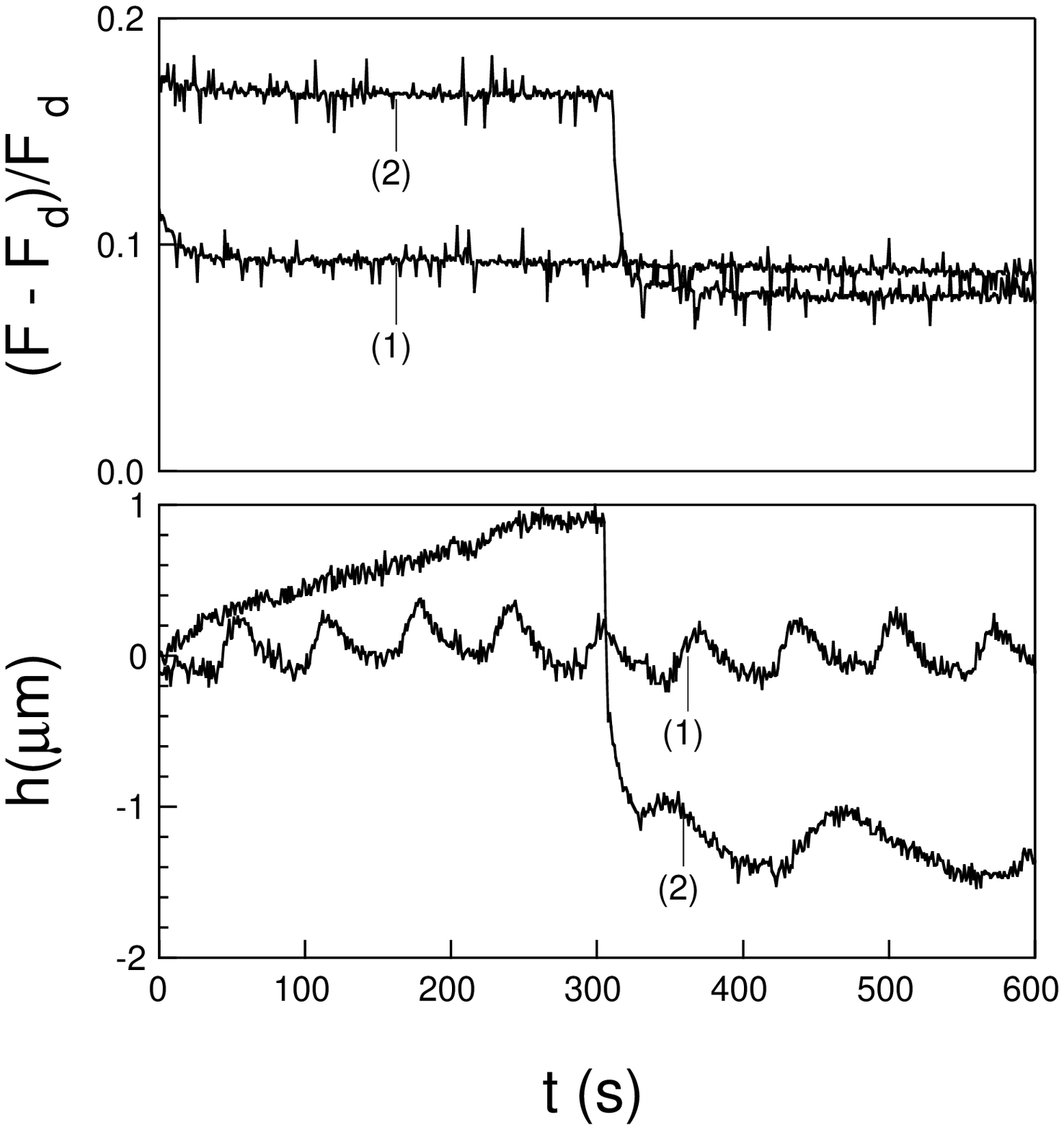,width=18pc,height=18pc,angle=0}
}
\caption{(a) Applied stress $F$ and (b) vertical position $h$ as functions
of time when a static stress larger than the critical value is applied
($k = 189.5~\rm N/m$, $M = 14.5~\rm g$, $V = 0$).  Two cases are shown:
(1) $F\simeq 1.1~F_d$; (2) $F \simeq 1.17~F_d$.}
\end{figure}

\begin{figure}[h]
\centerline{
\epsfig{file=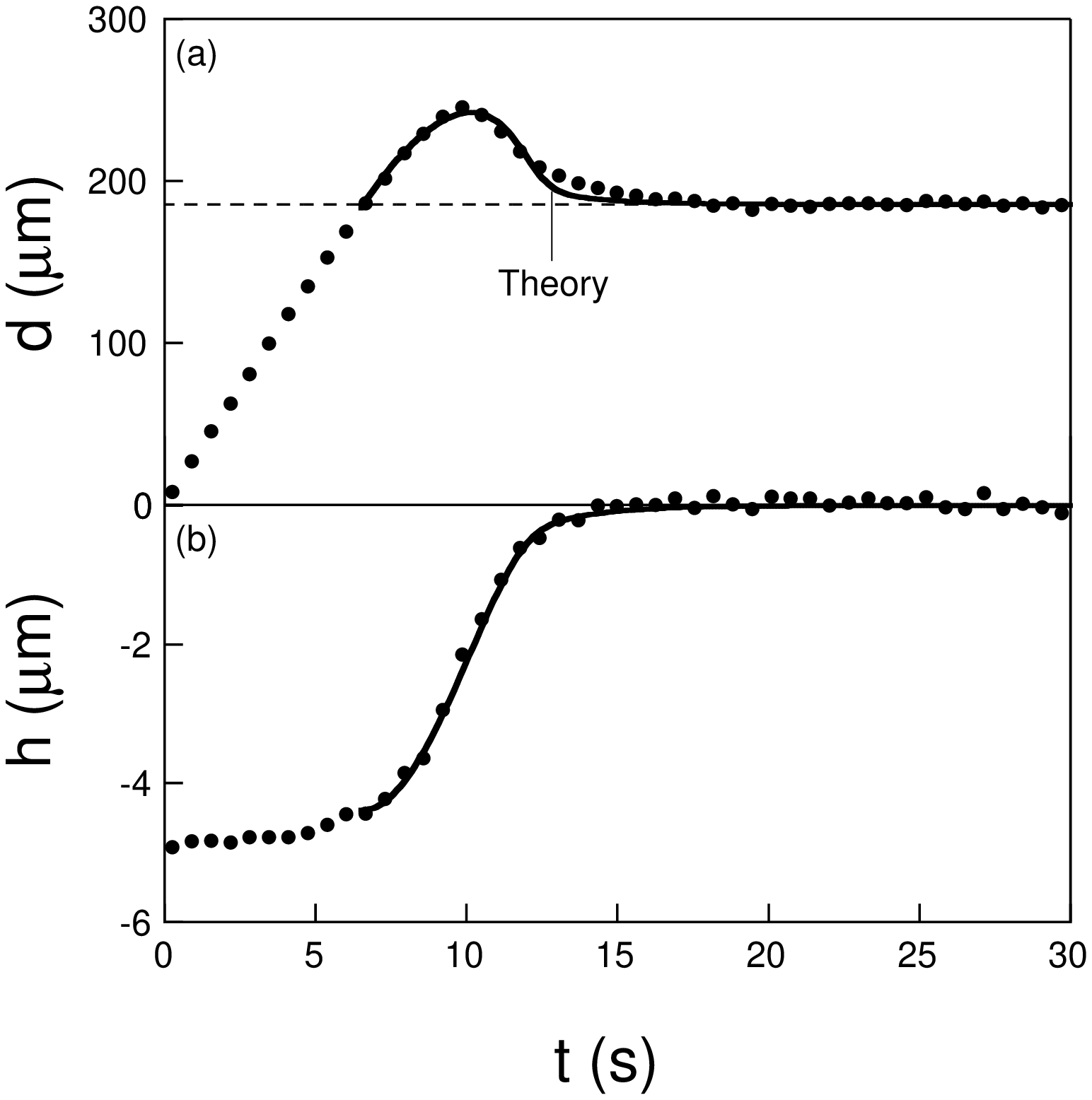,width=18pc,height=18pc,angle=0}
}
\caption{Experimental behavior compared with the theoretical model
(a) of the spring displacement $d(t)$ and
(b) of the vertical position $h(t)$ as a function
of time $t$ ($k = 189.5~\rm N/m$, $M = 14.5~\rm g$,
$V = 28.17~ \mu \rm m/s$). dots - experimental data.
lines - empirical model.}
\end{figure}


\begin{thebibliography}{99}



\bibitem{Gollub} S. Nasuno, A. Kudrolli, A. Bak, and J. P. Gollub,
Phys. Rev. E {\bf 58},  2161 (1998).

\bibitem{Nasuno} S. Nasuno, A. Kudrolli, and J. P. Gollub,
Phys. Rev. Lett. {\bf 79}, 949 (1997).
 
\bibitem{Marone} C. Marone,  Annu. Rev. Earth Planet Sci. 
{\bf 26}, 643 (1998).

\bibitem{Scholz} C. H. Scholz, 
{\it The mechanics of earthquakes and faulting},
Cambridge, Cambridge Univ. Press (1990).

\bibitem{Persson} B. N. J. Persson,
{\it Sliding friction : physical principles and applications}, Springer, 
New York (1998).

\bibitem{Ciliberto} L. Bocquet, E. Charlaix, S. Ciliberto, and
J. Crassous,  Nature (London) {\bf 396}, 735 (1998).

\bibitem{Marone90} C. Marone, C. B. Raleigh, and C. H. Scholz 
J. Geophys. Res. {\bf 95} 7007 (1990); C. Marone, PAGEOPH {\bf 137},
409 (1991).

\bibitem{Perrin} F. Heslot, T. Baumberger, B. Perrin, B. Caroli, 
and C. Caroli, Phys. Rev. E {\bf 49}, 4973 (1994).

\bibitem{Baumberger} T. Baumberger, F. Heslot and B. Perrin
 Nature (London) {\bf 367}, 544 (1994).

\bibitem{Brady93} J.F. Brady, J. Chem. Phys. {\bf 99}, 567 (1993).

\bibitem{explanation} This {\it stick-slip} motion of the plate
occurs at very slow velocities (typically $0.1~\mu \rm m/s$),
and is the mark of a velocity weakening of the granular
layer for small velocity.
A detailed study of this regime is in principle possible
but would require a very long experimental effort.  Here we 
focus on the continuous motion and on the transient regime
at larger velocities. 

\bibitem{num_rec} W. H. Press, S. A. Teukolsky, W. T. Vetterling, and 
B. P. Flannery,
{\it Numerical Recipes}, Cambridge, Cambridge Univ. Press (1992).

\bibitem{Rice} J.R. Rice and J. Gu,  Pure Appl. Geophys. {\bf 121}, 
187 (1983).

\bibitem{Geminard} J.-C. G\'eminard, W. Losert, S. Nasuno, and J.P. Gollub, 
in preparation.
\end{thebibliography}
\end{document}